\documentclass[fleqn,usenatbib]{mnras}

\usepackage{newtxtext,newtxmath}

\usepackage[T1]{fontenc}
\usepackage{ae,aecompl}
\pdfoutput=1

\usepackage{graphicx}	
\usepackage{amsmath}	
\usepackage{amssymb}	
\usepackage{enumitem}   

\title[Constraining the Low-Mass IMF]{The statistical challenge of constraining the low-mass IMF in Local Group dwarf galaxies}

\author[El-Badry et al.]{
Kareem El-Badry,$^{1}$\thanks{E-mail: kelbadry@berkeley.edu}
Daniel R. Weisz,$^{1}$ and Eliot Quataert$^{1}$
\\
$^{1}$Department of Astronomy, University of California, Berkeley, CA, USA.\\}

\date{Accepted XXX. Received YYY; in original form ZZZ}

\pubyear{2017}

\begin{document}
\label{firstpage}
\pagerange{\pageref{firstpage}--\pageref{lastpage}}
\maketitle

\begin{abstract}
We use Monte Carlo simulations to explore the statistical challenges of constraining the characteristic mass ($m_c$) and width ($\sigma$) of a lognormal sub-solar initial mass function (IMF) in Local Group dwarf galaxies using direct star counts. 
For a typical Milky Way (MW) satellite ($M_{V} = -8$), jointly constraining $m_c$ and $\sigma$ to a precision of $\lesssim 20\%$ requires that observations be complete to $\lesssim 0.2 M_{\odot}$, if the IMF is similar to the MW IMF. A similar statistical precision can be obtained if observations are only complete down to $0.4M_{\odot}$, but this requires measurement of nearly 100$\times$ more stars, and thus, a significantly more massive satellite ($M_{V} \sim -12$). 
In the absence of sufficiently deep data to constrain the low-mass turnover, it is common practice to fit a single-sloped power law to the low-mass IMF, or to fit $m_c$ for a lognormal while holding $\sigma$ fixed. 
We show that the former approximation leads to best-fit power law slopes that vary with the mass range observed and can largely explain existing claims of low-mass IMF variations in MW satellites, even if satellite galaxies have the same IMF as the MW. 
In addition, fixing $\sigma$ during fitting leads to substantially underestimated uncertainties in the recovered value of $m_c$ (by a factor of $\sim 4$ for typical observations). 
If the IMFs of nearby dwarf galaxies are lognormal and \textit{do} vary, observations must reach down to $\sim m_c$ in order to robustly detect these variations. 
The high-sensitivity, near-infrared capabilities of JWST and WFIRST have the potential to dramatically improve constraints on the low-mass IMF. We present an efficient observational strategy for using these facilities to measure the IMFs of Local Group dwarf galaxies.

\end{abstract}




\section{Introduction}
A persistent uncertainty in extragalactic astronomy is whether the stellar initial mass function (IMF) varies with environment and/or cosmic time. Many theoretical models of star formation hold that the shape of the IMF is set by local variables such as the Mach number in molecular clouds \citep[e.g.][]{Padoan_1997, Hopkins_2013}, the local Jeans length \citep[e.g.][]{Larson_1998, Narayanan_2012}, or the gas pressure in star-forming regions \citep[e.g.][]{Krumholz_2011, Krumholz_2016}, and thus predict that the IMF for a particular star formation event should depend to some extent on the local gas density and metallicity, and on the instantaneous star formation rate \citep[see][for a review]{Offner_2014}.

On the other hand, most observational studies have measured mass functions consistent with a universal IMF across a wide variety of environments \citep[see][for a review]{Bastian_2010}. Some works \textit{have} found integrated light signatures of low-mass IMF variations in the centers of massive early-type galaxies \citep{vanDokkum_2008, Conroy_2012, Cappellari_2012, Smith_2014, vanDokkum_2016}, but these results require further interpretation \citep{McConnell_2016, Coulter_2016}. Most IMF studies carried out with direct star counts have found little evidence of IMF variation in the Local Group \citep{Bessell_1993, Wyse_2002, Chabrier_2003, Covey_2008, Bastian_2010, Bochanski_2010}. 

Dwarf spheroidal and ultra-faint galaxies in the Local Group provide a natural laboratory in which to search for environmental variations in the IMF. Compared to the Milky Way (MW), Local Group satellites have extremely low stellar densities, low metallicities ($\rm [Fe/H]\sim -2$), high alpha abundances ($\rm [\alpha/Fe]\sim 0.4$), and high mass-to-light ratios ($M/L\sim 100$, in solar units) \citep{Tolstoy_2009, Frebel_2015}. In addition, many (though not all) Local Group dwarf galaxies exhibit relatively simple star formation histories, having formed all their stars within $\lesssim 2$ Gyr, before $z \sim 3$ \citep{Brown_2012, Vargas_2013, Webster_2015, Weisz_2014}. One might thus expect systematic IMF variation between Local Group dwarf galaxies and the MW. 

Observational searches for IMF variations in MW satellites have thus far yielded mixed results: \citet{Grillmair_1998} and \citet{Wyse_2002} found IMFs in the Draco and Ursa Minor dwarf spheroidal galaxies that were roughly consistent with the IMF measured in the MW and in its globular clusters, and \citet{Kalirai_2013} found the IMF of the Small Magellanic Cloud (SMC) to be only marginally shallower than that of the MW. On the other hand, \citet{Geha_2013} found the IMFs of the Hercules and Leo IV ultra-faint dwarf galaxies to be significantly shallower (more bottom-light) than the canonical MW IMF. 

All published attempts to measure the IMFs of nearby dwarf galaxies have been limited at low masses by observational sensitivity. At a distance of $D = 150$ kpc, Hubble Space Telescope (HST) photometric studies become incomplete below stellar masses of $\sim 0.5\,M_{\odot}$. This is significantly more massive than the characteristic mass at which the canonical MW IMF turns over and begins to significantly deviate from a power law ($m_{c} \sim 0.22 M_{\odot}$). No published observational studies have detected a turnover in the mass functions of nearby dwarf galaxies, and they have thus all opted to fit a simple power law IMF. However, measuring the turnover in the IMF (if indeed it exists) is particularly desirable, as it would provide a direct test of theoretically predicted scalings of $m_c$ with metallicity and density. 

The statistical power of resolved stellar population studies to constrain the low-mass IMF will improve significantly with the introduction of the James Webb Space Telescope (JWST), whose larger collecting area and improved near-infrared sensitivity will allow observations to probe the IMFs of Local Group galaxies to lower masses than past HST observations. In the nearest MW satellites, JWST will efficiently observe down to the hydrogen burning limit, making it possible to constrain the characteristic mass and slope of the IMF with unprecedented precision.   

In this paper, we use Monte Carlo simulations to investigate how the accuracy to which the IMF can be recovered from star counts scales with the size and limiting magnitude of the observed stellar sample. We demonstrate that HST resolved stellar population studies do not have the statistical power to robustly detect IMF variations in most Local Group dwarf galaxies if variation occurs primarily at lower masses. However, we find that similar studies with JWST will have the potential to definitively quantify (or rule out) low-mass IMF variations for galaxies within $\sim 100$ kpc. 

We organize this paper as follows.  In Section~\ref{sec:methods}, we describe how we fit the IMF from a sample of masses or a color-magnitude diagram (CMD). In Section~\ref{sec:recovering_true_imf}, we explore how accurately the parameters of a MW-like lognormal IMF can be constrained with observations reaching a range of limiting magnitudes. We then investigate the systematics introduced by simplifications which are commonly made in the absence of sufficiently deep observations, such as fitting a single power law or holding one of the parameters of a lognormal fixed during fitting. In Section~\ref{sec:true_variation}, we investigate what observations are required to detect variation in the IMF. In Section~\ref{sec:looking_forward}, we discuss how future observations with JWST will improve the constraints on the IMFs of nearby dwarf galaxies, and we lay out an efficient observing strategy for characterizing IMF variations in the Local Group with JWST. We summarize our results in Section~\ref{sec:summary}.

\section{Methods}
\label{sec:methods}

\subsection{Lognormal IMF}
\label{sec:method_lognormal_imf}

\begin{figure}
\includegraphics[width=\columnwidth]{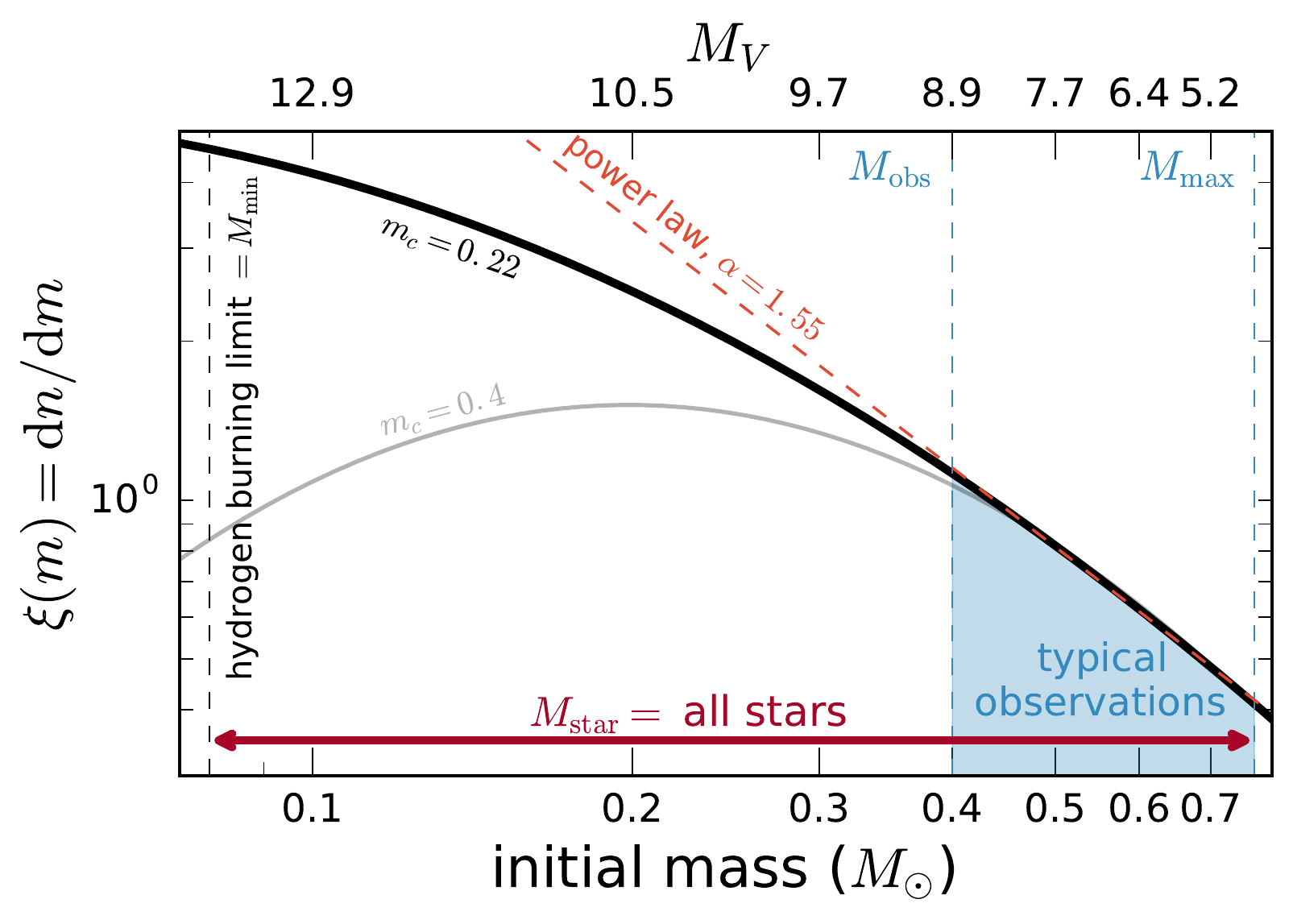}
\caption{Schematic illustration of the challenge of constraining the low-mass IMF. Observations sample the IMF between $M_{\rm obs}$, the lowest stellar mass which can be reached by observations, and $M_{\rm max}$, the highest stellar mass which remains on the main sequence. The total mass $M_{\rm star}$ corresponds to all stars between $M_{\rm min}$ and $M_{\rm max}$, many of which cannot be observed. When observations do not reach the low-mass turnover, both power law IMFs and lognormal IMFs with a wide range of characteristic masses $m_c$ are consistent with the data.}
\label{fig:overview}
\end{figure}

We assume that the true IMF from which masses are sampled is a lognormal of the form advocated by \citet{Chabrier_2003ApJ}, which at sub-solar masses is given by
\begin{equation}
\label{eqn:chab_imf}
\xi\left(m\right)=\frac{{\rm d}n}{{\rm d}m}=\xi_{0}\frac{1}{m}\exp\left[-\frac{\left(\log m-\log m_c \right)^{2}}{2\sigma^{2}}\right],
\end{equation}
where the logarithm is base 10. Here $m_c = 0.22 M_{\odot}$ is the ``characteristic mass'' at which the logarithmic slope of the IMF is ${\rm d}\log \xi/{\rm d} \log m= -1$,\footnote{Note that $m_c$ is \textit{not} the most probable mass or the peak of the mass probability distribution function ${\rm d}n/{\rm d}m$. The peak is at $m_{c}\exp\{-\left[\ln\left(10\right)\sigma\right]^{2}\}$, which is always less than $m_c$. $m_c$ is the peak of ${\rm d}n/{\rm d}\log m,$ the probability density function in log space.} and $\sigma = 0.57 M_{\odot}$.\footnote{These parameters are for the ``system'' IMF, which represents the observable mass function after the single-star IMF is convolved with the distribution of unresolved multiple-star systems.} Hereafter, we refer to a lognormal IMF with these parameter values as a ``Chabrier IMF.'' The normalization factor $\xi_0$ is defined such that $\int_{M_{{\rm min}}}^{M_{{\rm max}}}\xi\left(m\right)\,{\rm d}m=1$, where $M_{\rm min}$ and $M_{\rm max}$ are the minimum and maximum of the mass range over which the IMF is populated. For a single-age stellar population, $M_{\rm max}$ is the highest initial mass which has not yet left the main sequence, and $M_{\rm min}$ is taken to be the hydrogen burning limit, $M_{\rm min}\approx 0.08 M_{\odot}$ \citep[e.g.,][]{Kumar_1963, Grossman_1970, Liebert_1987}. Throughout this work, we take $M_{\rm max}=0.77 M_{\odot}$, which is approximately the highest mass in a stellar population with $\rm Fe/H =-2$ that remains on the main sequence after 12.5 Gyr \citep{Paxton_2011, Choi_2016}.

Figure~\ref{fig:overview} illustrates the typical situation for observational IMF studies of Local Group dwarf spheroidal and ultra-faint galaxies. Since these galaxies are dominated by old stellar populations, they have no surviving stars above $M_{\rm max} \approx 0.77 M_{\odot}$, and because low-mass stars are faint, observations typically do not reach the low-mass turnover. Published observational studies of Local Group dwarf galaxies thus far have, at best, reliably sampled the IMF down to stellar masses of $M_{\rm obs}\sim 0.4 M_{\odot}$, for observations of very near satellites \citep[$D = 60-70$ kpc;][]{Wyse_2002,Kalirai_2013}. Studies of more distant satellites (especially older studies using WFPC2) have only reached down to $M_{\rm obs} = 0.5 - 0.6 M_{\odot}$ \citep{Grillmair_1998, Geha_2013}. 

If observations do not reach down to the low-mass turnover, one is faced with the underconstrained problem of inferring the form of the IMF from observations sampling a narrow mass range in which the IMF has a near constant slope. 
As Figure~\ref{fig:overview} illustrates, the Chabrier IMF is effectively indistinguishable from a power law ($\xi(m)\propto m^{-\alpha}$, where $\alpha=2.35$ is the Salpeter value) with $\alpha \approx 1.55$ in the mass range $(0.4 - 0.77) M_{\odot}$. Data drawn from a Chabrier IMF in this mass range are also consistent with having been drawn from a wide range of lognormal IMFs with different characteristic masses $m_c$ (and appropriately chosen $\sigma$ values; e.g., $\sigma = 0.57M_{\odot}$ for $m_c = 0.22M_{\odot}$, or $\sigma = 0.36M_{\odot}$ for $m_c = 0.4M_{\odot}$). In general, if both $m_c$ and $\sigma$ are allowed to vary, observations which reach a lower mass limit $M_{\rm obs}$ and sample a lognormal IMF will be consistent with any lognormal IMF with $m_c \in [0, M_{\rm obs}]$. 

\subsection{Sampling from the IMF}
We draw masses from a lognormal IMF using inverse transform sampling, as described in Appendix~\ref{sec:drawing_masses}. In practice, observations cannot directly recover the true single-star IMF, even if observations are arbitrarily deep and can measure stellar masses with zero uncertainty. This is because a fraction of stars are in unresolved multiple-star systems (e.g., binaries) and thus appear in observations as a single star with the combined magnitude of all components, corresponding to a higher inferred mass. This effect will make the observationally inferred IMF slightly more top-heavy (i.e., steeper) than the underlying single-star IMF, though this effect only becomes significant at $m\lesssim 0.3 M_{\odot}$ \citep{Kroupa_1991}.

Rather than explicitly modeling unresolved multiple-star systems, which have poorly constrained binary fractions and mass ratio distributions, we draw masses from the ``system'' IMF reported in \citet{Chabrier_2003ApJ}. The system IMF is =observationally motivated: it represents the IMF derived from the system luminosity function, which is produced by ``merging'' the components of each multiple-star system into a single unresolved source.\footnote{Note that this is defined differently from e.g. the system IMF of \citet{Kroupa_2013}, which describes the distribution of the total masses of multiple-star systems. The Chabrier system IMF is derived by applying a single-star mass-magnitude relationship to an unresolved multiple-star system.} The tests we carry out here thus quantify the accuracy with which the observable system IMF can be recovered; additional modeling is required to translate the system IMF into a single-star IMF.

\subsection{Fitting the IMF}
\label{sec:fitting}
Recovering the low-mass IMF from an observed sample of stars is straightforward compared to the high-mass IMF, because to first order, the galaxy's star formation history (SFH) is irrelevant \citep{Miller_1979, Scalo_1986}. Since the main-sequence lifetimes of low-mass ($\lesssim 0.75 M_{\odot}$) stars exceed the age of the universe, the present-day mass function is identical to the IMF at low masses, irrespective of the galaxy's star formation history.\footnote{Strictly speaking, SFH \textit{can} have a slight effect on the recovered IMF, because there is weak evolution in the magnitudes of low-mass stars even while they remain on the main sequence. However, this effect introduces negligible uncertainty in the recovered low-mass IMF \citep{Geha_2013}, because the main-sequence lifetimes of low-mass stars are many times the age of the Universe.}

We experimented with two different methods for fitting the IMF to observations. The first method is idealized, with the assumption that the masses of individual stars can be measured exactly. In this case, the probability of measuring a particular mass is simply the value of the IMF at that mass. Assuming masses are independent and identically distributed, the likelihood function for a set of stellar masses is simply the product of the probability of each mass. Explicitly, given a sample of masses $m_i$ and a functional form of the IMF $\xi(m, \theta_j)$, where $\theta_j$ are free parameters, the likelihood function is \begin{equation}
\label{eqn:likelihood}
\mathcal{L}=p\left(m_{i}|\theta_{j}\right)=\prod_{i}\xi\left(m_{i},\theta_j\right).
\end{equation}
An advantage of this approach compared to the traditional method of essentially fitting a line to a histogram is that it does not require any binning. Binning masses when fitting a distribution introduces unnecessary ambiguity in the choice of bins and can lead to biases in the inferred distribution, particularly when there are unequal numbers of samples in each bin \citep[e.g.][]{MaizApellaniz_2005, Maschberger_2009}.  

We also experimented with a more realistic (though still idealized), observationally motivated fitting approach using synthetic CMDs generated for a stellar population with a realistic completeness and metallicity distribution function. In this case, it is not possible to determine the mass of any individual star exactly, since stars of different metallicities have different magnitudes at fixed initial mass. 

We describe this approach in detail and compare the results of CMD fitting to the idealized approach of fitting masses directly in Appendix~\ref{sec:cmd_fitting}. We find that, if the metallicity distribution function and observational completeness are known, the two fitting approaches yield effectively identical best-fit values, covariances, and marginalized uncertainties in IMF parameters. That is, in the limit of zero photometric uncertainty and perfect stellar models, no information is lost by fitting observables as opposed to fitting masses directly as in Equation ~\ref{eqn:likelihood}. Having demonstrated as a proof of concept that both fitting approaches produce the same results in Appendix~\ref{sec:cmd_fitting}, we use the first approach (directly fitting masses) exclusively for our primary analysis in the interest of computational cost.  

For both fitting approaches, we assume flat priors for the free parameters of the IMF. When fitting a lognormal, we use the priors $m_c \in [0.01, 1.5]\times M_{\odot}$ and $\sigma\in[0.1, 2] \times M_{\odot}$; when fitting a power law, we take $\alpha\in[0, 5]$. We have verified that our priors are wide enough to be noninformative; that is, increasing the range of the priors does not change the shape of the recovered posterior. 

We use the affine-invariant ensemble sampler \texttt{emcee}  \citep{ForemanMackey_2013} to sample from the posterior. To verify that our chains have converged, we use the Gelman-Rubin convergence diagnostic \citep{Gelman_1991, Gelman_2013}. Formally, we require $\hat{R}<1.05$ for all chains, where $\hat{R}$ is the potential scale reduction factor (PSRF).\footnote{The PSRF quantifies the ratio of the mean variance within individual chains to the variance of the mean across all chains. In the limit of infinite samples, $\hat{R}\to 1$; values of $\hat{R}$ greater than $\sim 1.1$ indicate poor convergence \citep{Gelman_2013}.} We find that in practice, drawing 20,000 samples from the posterior is always sufficient to satisfy this diagnostic threshold. 

\section{Recovering the parameters of a Lognormal IMF}
\label{sec:recovering_true_imf}
In this section, we investigate how accurately the parameters of a lognormal IMF can be recovered from a sample of masses. We begin by exploring how the observational mass lower-limit $M_{\rm obs}$ and total stellar mass $M_{\rm star}$ of a sample affect the strength of constraints on $m_c$ and $\sigma$, as well as the degeneracy between the two parameters. We assume that the true IMF is a Chabrier lognormal IMF with $m_c = 0.22 M_{\odot}$ and $\sigma = 0.57 M_{\odot}$. We consider the case in which the true IMF \textit{does} vary in Section~\ref{sec:true_variation}. 

\subsection{Overview}
\label{sec:overview}
We simulate observations of a galaxy with total stellar mass $M_{\rm star}$ and a Chabrier IMF between $M_{\rm min} = 0.08M_{\odot}$ and $M_{\rm max} = 0.77 M_{\odot}$ by drawing $N_{\rm obs}$ masses from the IMF between $M_{\rm obs}$ and $M_{\rm max}$.\footnote{In this formulation, $M_{\rm star}$ represents the total stellar mass in the region of the galaxy covered by observations. If only a fraction of the galaxy is observed, the total stellar mass will be greater than $M_{\rm star}$. Note also that $M_{\rm star}$ represents only the initial mass contribution of main-sequence stars. We do not account for mass loss, evolved stars, or remnants.} Due to incompleteness at lower masses, only the fraction of stars $c_{\rm obs}$ with masses greater than $M_{\rm obs}$ can be observed, so 

\begin{equation}
\label{eqn:N_obs}
N_{{\rm obs}}=N_{{\rm tot}}c_{{\rm obs}}=\frac{M_{{\rm star}}}{\overline{m}}c_{\rm obs}=M_{{\rm star}}\frac{\int_{M_{{\rm obs}}}^{M_{{\rm max}}}\xi\left(m\right)\,{\rm d}m}{\int_{M_{{\rm min}}}^{M_{{\rm max}}}\xi\left(m\right)m\,{\rm d}m}.
\end{equation}

In practice, the lower-limit for observations is not a threshold mass, but rather a magnitude threshold below which completeness falls off steeply because stars are too faint to observe or observations become crowding limited. We translate magnitude thresholds in a given filter to $M_{\rm obs}$ by interpolating on a grid of isochrones from the MESA Isochrones and Stellar Tracks project \citep[\textsc{Mist;}][]{Choi_2016}, as explained in Appendix~\ref{sec:stellar_models}. For this conversion, we assume a stellar population of age 12.5 Gyr with metallicity $\rm [Fe/H]=-2$. 

\subsection{Fitting results}
\label{sec:recovering_c03}

\begin{figure*}
\includegraphics[width=\textwidth]{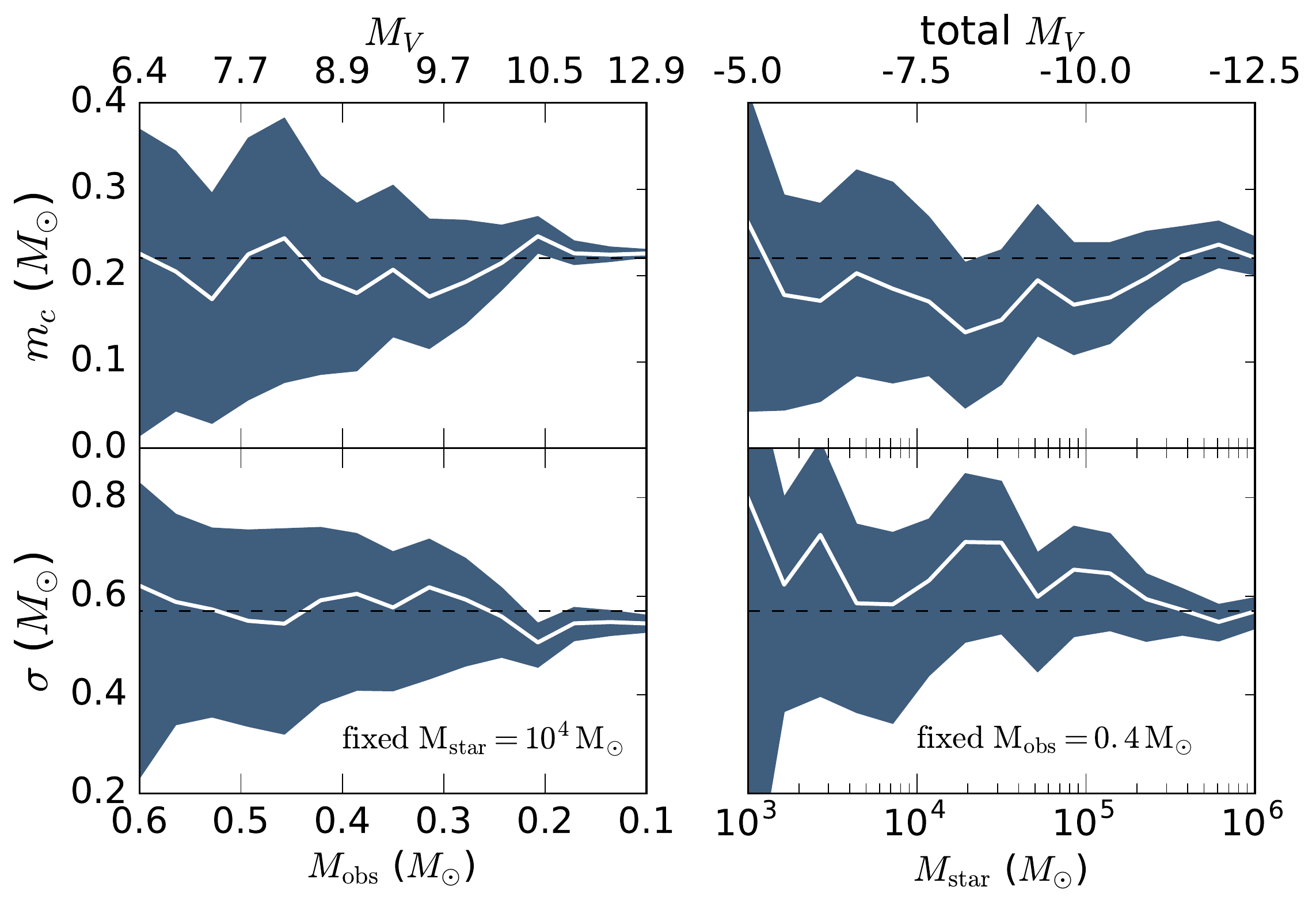}
\caption{Best-fit values of $m_c$ (top) and $\sigma$ (bottom) obtained from fitting a lognormal IMF to samples of stars drawn from a Chabrier IMF. White line and shaded regions show the median and middle 68\% of the maginalized posterior for each IMF parameter. \textbf{Left}: we fix the total mass of all stars in the range $0.08-0.77\,M_{\odot}$ at $10^{4}\,M_{\odot}$ and vary $M_{\rm obs}$, the mass threshold down to which stars can be observed. \textbf{Right}: we fix $M_{\rm obs}$ at $0.4\,M_{\odot}$ (corresponding to a magnitude threshold of $M_V \approx 8.9$) and vary the total stellar mass. Tight constraints on $m_c$ and $\sigma$ can be obtained either by observing systems with more stars or by pushing observations to fainter magnitudes, but unrealistically large stellar samples are needed to meaningfully constrain the low-mass the IMF at the magnitude limits of current studies with HST.}
\label{fig:sample_size_and_limiting_mag}
\end{figure*}

Figure~\ref{fig:sample_size_and_limiting_mag} shows the constraints on $m_c$ and $\sigma$ obtained by fitting a lognormal IMF to a range of mass samples drawn from a Chabrier IMF. The white lines and blue shaded regions respectively show the median and middle 68 percent of the marginalized posterior distributions for $m_c$ (top) and $\sigma$ (bottom). Large shaded regions indicate significant uncertainty in the IMF parameters, due both to the degeneracy between $m_c$ and $\sigma$ and to stochastic sampling effects arising from the finite number of stars being sampled \citep{Elmegreen_1999, Hernandez_2012}. Smaller regions indicate that $m_c$ and $\sigma$ are individually well constrained. 

In the left two panels, we generate a stellar population with  total mass $M_{\rm star} = 10^4 M_{\odot}$ (corresponding to $N_{\rm tot} = 34,287$ between $0.08 M_{\odot}$ and $0.77 M_{\odot}$) and vary $M_{\rm obs}$, the mass down to which stars can be observed. At $M_{\rm obs} = 0.4 M_{\odot}$, this $M_{\rm star}$ corresponds to $N_{\rm obs} = 8,686$ stars included in the fit, which is a factor of a few more than the number used in most previous observational studies of Local Group dwarf galaxies. 

For $M_{\rm obs} \gtrsim 0.2 M_{\odot}$, the middle 68\% ranges for both $m_c$ and $\sigma$ are large, indicating weak constraints on the individual IMF parameters. As we will show, this is largely the result of the degeneracy between $m_c$ and $\sigma$. Strong constraints on $m_c$ and $\sigma$ can be obtained only when $M_{\rm obs} \lesssim m_c$. This makes intuitive sense, since the lognormal IMF has a nearly constant slope at $m\gg m_c$ and only begins to turn over at $m\sim m_c$. Above the turnover, $m_c$ and $\sigma$ are strongly degenerate, because a wide range of lognormal IMFs with $0 \lesssim m_c \lesssim M_{\rm obs}$ can all fit the data equally well (see Figure~\ref{fig:overview}). 

In the right panels of Figure~\ref{fig:sample_size_and_limiting_mag}, we again show constraints on $m_c$ and $\sigma$, but now vary $M_{\rm star}$ while $M_{\rm obs}=0.4M_{\odot}$ is held fixed. As expected, stronger constraints can also be obtained by increasing $M_{\rm star}$, but very large numbers of stars are required to obtain meaningful constraints. For example, $M_{\rm star}$ must be increased by a factor of $\sim 10$ to obtain the same improvement on the fiducial constraints that can be obtained by decreasing $M_{\rm obs}$ by $0.2 M_{\odot}$. 

For reference, Figure~\ref{fig:sample_size_and_limiting_mag} also shows the absolute magnitudes of individual stars (left) and the total galaxy (right) that correspond to the plotted values of $M_{\rm obs}$ and $M_{\rm star}$. We calculate $M_V$ for individual stars from the \textsc{Mist} isochrones as explained in Appendix~\ref{sec:stellar_models} and $M_V$ for a whole galaxy using Equation~\ref{eqn:mstar_mv}.

\subsection{Fitting a power law IMF to data drawn from a lognormal}
\label{sec:fixing_params_danger}

\begin{figure}
\includegraphics[width=\columnwidth]{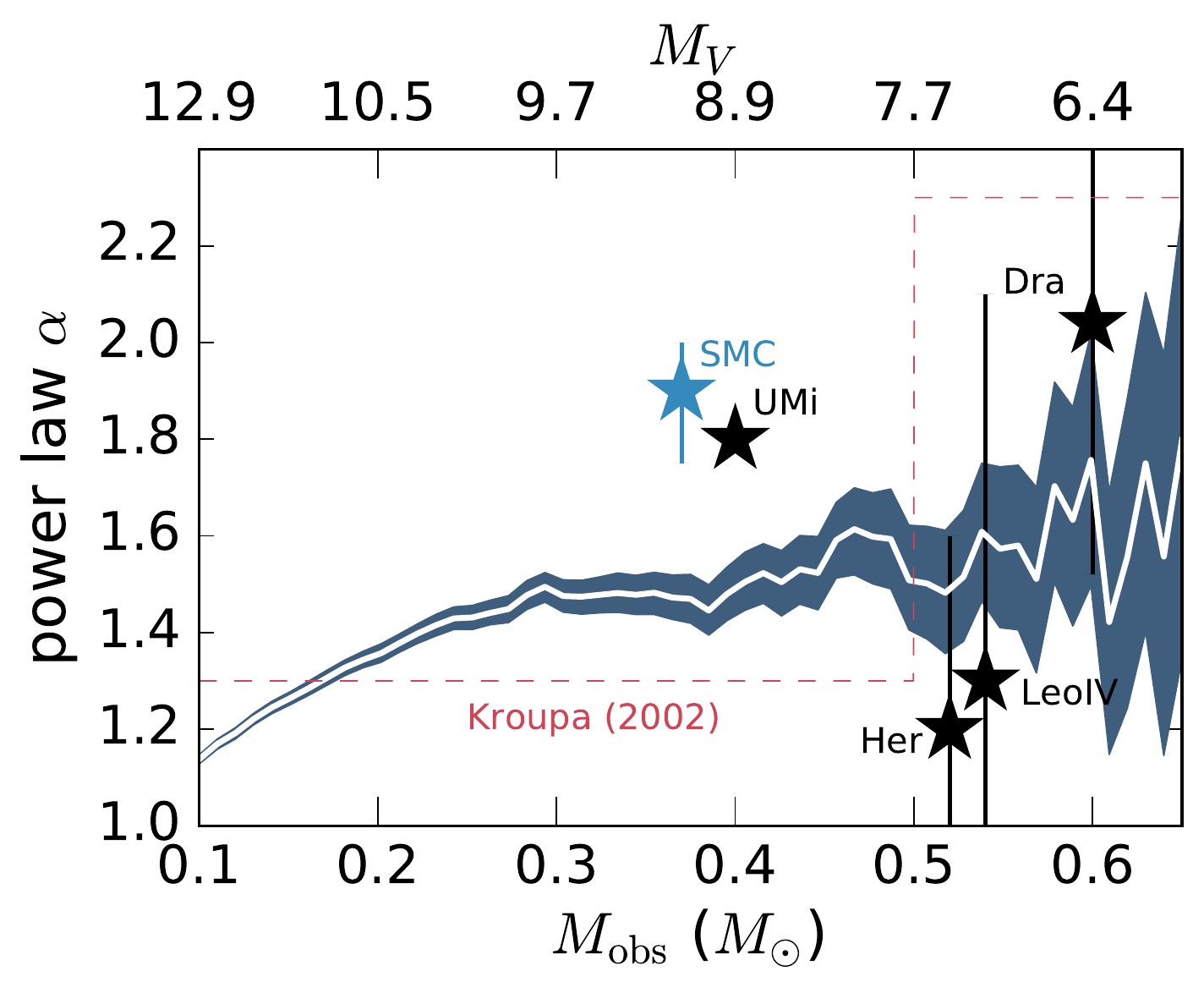}
\caption{Best fit logarithmic IMF slope ($\xi(m)\propto m^{-\alpha}$) obtained by fitting a power law to masses drawn from a Chabrier lognormal IMF (with total $M_{\rm star}=10^4 M_{\odot}$) between $M_{\rm obs}$ and $M_{\rm max} = 0.77 M_{\odot}$. Fitting a power law to observations which reach $M_{\rm obs}\sim 0.5 M_{\odot}$ yields a logaritmic slope $\alpha \sim 1.55$, which is significantly shallower than than the canonical value of $\alpha = 2.3$. Observationally-inferred IMF constraints are overplotted for five Local Group dwarf galaxies: Ursa Minor \citep[][no error bars included]{Wyse_2002}, Hercules and Leo IV \citep{Geha_2013}, Draco \citep{Grillmair_1998}, and the Small Magellanic Cloud \citep[SMC;][]{Kalirai_2013}. Note that the first four galaxies are dominated by old stellar populations, while the SMC sample contained some younger stars.}
\label{fig:alpha_vs_mobs}
\end{figure}

In the absence of sufficiently deep observations with the statistical power to jointly constrain $m_c$ and $\sigma$ of a lognormal IMF, it is common to make simplifying assumptions, such as fitting a single power law in place of a lognormal or holding $\sigma$ fixed when fitting a lognormal \citep[e.g.][]{Grillmair_1998, Wyse_2002, Geha_2013, Kalirai_2013}. This avoids the degeneracy between $m_c$ and $\sigma$, since either approach essentially reduces the fit to a one dimensional optimization problem. 

Figure~\ref{fig:alpha_vs_mobs} shows the results of fitting a power law ($\xi(m)\propto m^{-\alpha}$) to masses drawn from a Chabrier IMF. As in the left panel of Figure~\ref{fig:sample_size_and_limiting_mag}, we hold $M_{\rm star}$ fixed at $10^4 M_{\odot}$ while varying $M_{\rm obs}$. Here, the 68 percent probability range in $\alpha$ is narrow: for $M_{\rm obs} \lesssim 0.5 M_{\odot}$, $\alpha$ is constrained within $\lesssim 10$\%. The improved constraints on $\alpha$ at lower $M_{\rm obs}$ primarily reflect the fact that deeper observations increase the total number of stars included in the fit. 

The best-fit logarithmic slope $\alpha$ becomes shallower for decreasing $M_{\rm obs}$, since deeper observations begin to sample the mass regime near the turnover in the IMF. However, for values of $M_{\rm obs}$ in the range $(0.3 - 0.6)M_{\odot}$ (with $M_{\rm max} = 0.77 M_{\odot}$), the best-fit logarithmic slope is almost constant at $\alpha \approx 1.55$. This suggests that, for this choice of $M_{\rm star} = 10^4 M_{\odot}$, observations would have to reach $M_{\rm obs} \lesssim 0.3 M_{\odot}$ before they could begin to detect deviations from a power law. 

Figure~\ref{fig:alpha_vs_mobs} also shows the results of previous studies which have fit power law IMFs to star count data from Local Group dwarf galaxies. Black stars show results for ultra-faint and dwarf spheroidal galaxies, which have very low metallicities, no gas, and are dominated by old stellar populations; we plot results for Draco \citep{Grillmair_1998}, Ursa Minor \citep{Wyse_2002}, Hercules and Leo IV \citep{Geha_2013}. We also show the result of \citet{Kalirai_2013} for the IMF of the SMC; we plot this point in blue to emphasize the SMC's more complicated star formation history. These previous studies have reached down to $M_{\rm obs}$ values between $0.37 M_{\odot}$ and $0.6 M_{\odot}$. None of them detected a turnover at lower masses, and they thus all opted to fit a single power law.

We note that the best-fit logarithmic slope of $\alpha \sim 1.55$ for masses in the range typically probed by observations ($0.4M_{\odot} \lesssim m \lesssim 0.77 M_{\odot}$) is significantly shallower than the \citet{Kroupa_2002} value of $\alpha = 2.3$ for $m > 0.5 M_{\odot}$ and the \citet{Salpeter_1955} value of $\alpha = 2.35$ at all masses. Previous studies have suggested that the shallower $\alpha$ values measured in Local Group dwarf galaxies are evidence of systemic variation in the the IMF with environment. For example, \citet{Geha_2013} pointed out that the value $\alpha = 1.2_{-0.5}^{+0.4}$ found in Hercules is shallower than the Kroupa value at the $5.4\sigma$ level. However, Figure~\ref{fig:alpha_vs_mobs} shows that the shallowest observationally-inferred IMFs are all approximately consistent with having been drawn from a Chabrier lognormal IMF.

There are two reasons for this somewhat counterintuitive result. First, the Chabrier and Kroupa IMFs, though similar in overall shape, have different instantaneous slopes: at $m=0.5 M_{\odot}$, the Chabrier IMF has an instantaneous logarithmic slope ${\rm d}\log \xi /{\rm d}\log m \approx -1.48$; at the same mass, the logarithmic slope of the Kroupa IMF steepens to $-2.3$.\footnote{Note, however, that it is not the instantaneous slope at $M_{\rm obs}$, but rather the average slope between $M_{\rm obs}$ and $M_{\rm max}$, that drives the fit.} Thus, slopes of $\alpha\sim 1.55$ at $m\geq 0.5M_{\odot}$ appear significantly shallower than the Kroupa value (the 99\% ($3 \sigma$) confidence interval on the Kroupa slope is $\alpha = 2.3 \pm 0.3$ in this mass range \citep{Kroupa_2001}), but are entirely consistent with being drawn from a Chabrier IMF. We therefore stress that, although many works treat both the Kroupa and Chabrier IMFs as identical ``canonical'' IMFs, the two IMFs make significantly different predictions for observations which sample only a narrow range of masses near $0.5 M_{\odot}$.

Second, the logarithmic slope of the Kroupa IMF steepens discontinuously from $\alpha = 1.3$ to $\alpha = 2.3$ at $m=0.5 M_{\odot}$. Thus, although the shallower IMF slopes found in Local Group dwarf galaxies thus far \textit{are} significantly different from the Kroupa value at $m\geq 0.5 M_{\odot}$, they are consistent with the Kroupa value at slightly lower masses. As there is little physical reason to expect a discontinuity in the slope of the IMF \citep{Miller_1979}, one might reasonably expect intermediate $\alpha$ values between $1.3$ and $2.3$ at masses near $0.5 M_{\odot}$. 

\citet{Kalirai_2013} interpreted their best-fit slope of $\alpha = 1.90_{-0.10}^{+0.15}$ as an indication that the SMC's IMF is somewhat shallower at lower masses than the Kroupa value, and significantly shallower than the $\alpha=2.35$ prediction from \citet{Salpeter_1955}. Figure~\ref{fig:alpha_vs_mobs} shows that the measured slope is actually somewhat \textit{steeper} than what would be measured if masses were drawn from a Chabrier IMF in the range $0.37 M_{\odot} \leq m \leq 0.77 M_{\odot}$. However, \citet{Kalirai_2013} fit masses up to $M_{\rm max} = 0.93M_{\odot}$ rather than $0.77 M_{\odot}$. Adopting their $M_{\rm max}$ and $M_{\rm min}$, we find $\alpha = 1.58^{+0.06}_{0.05}$ for masses drawn from a Chabrier IMF. This is slope is shallower than the SMC value, but only at the $(2-3)\sigma$ level.

Finally, we note that, with the exception of the data for Ursa Minor, the observational results plotted in Figure~\ref{fig:alpha_vs_mobs} were obtained in studies that attempted to explicitly model the effects of binaries and recover the single-star IMF, not the system IMF. The system IMFs corresponding to these results are expected to be slightly shallower (lower $\alpha$). However, as demonstrated by \citet[][their Figure 3]{Kroupa_1991}, accounting for binarity has negligible effects on the inferred IMF except at very low masses ($m \lesssim 0.3$). We find that at $m\geq 0.4 M_{\odot}$, the single-star and system IMFs reported in \citet{Chabrier_2003} differ by less than 10\%, so we do not expect binarity to significantly change our results. 

However, future observations with JWST will probe the IMFs of nearby dwarf galaxies to significantly lower masses (see Section~\ref{sec:looking_forward}), and we expect more significant differences between the system and single-star IMFs in this regime. In particular, because the conversion between the system and single-star IMFs depends on the assumed binary fraction, environmental variations in the binary fraction or mass ratios in binary systems could lead to variation in the system IMF, even in the absence of variation in the single-star IMF.

\subsection{Distinguishing a lognormal IMF from a power law}

\begin{figure}
\includegraphics[width=\columnwidth]{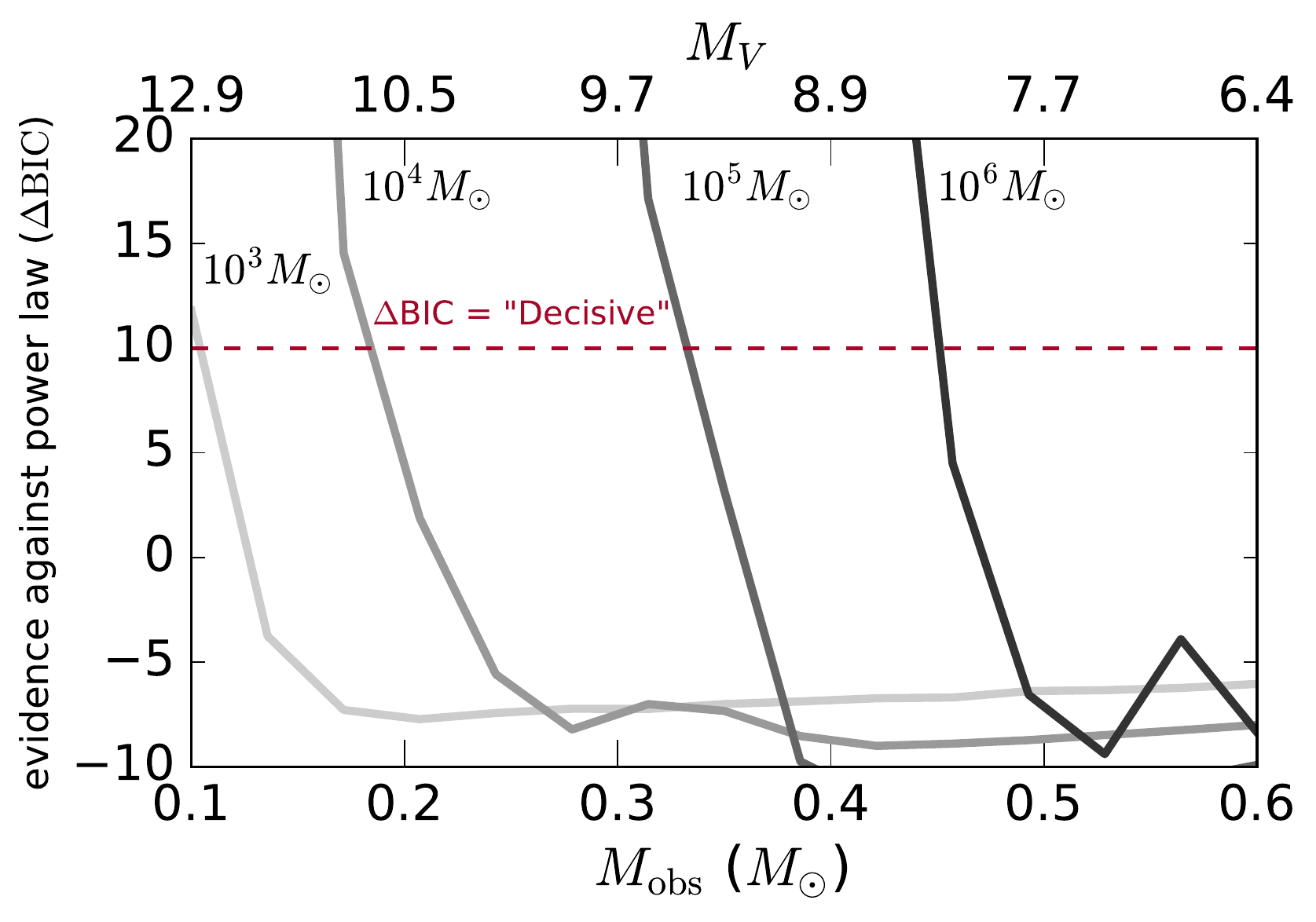}
\caption{Difference in Bayesian information criterion (BIC) between a power law IMF and a lognormal. Higher $\Delta$BIC indicates stronger deviations from a power law and evidence for a lognormal. In all cases, we draw masses from a Chabrier lognormal IMF. Then, using only masses above a threshold $M_{\rm obs}$, we fit both a power law and a lognormal to the sampled masses and calculate the BIC. Correctly inferring that the IMF is lognormal, not a power law, requires either deep observations (low $M_{\rm obs}$) or a very large sample (high $M_{\rm star}$).}
\label{fig:delta_bic}
\end{figure}

We now consider how deep observations must reach in order to robustly detect deviations from a power law if the true IMF is lognormal. Existing observational studies of Local Group dwarf galaxies have shown that their IMFs do not show any deviation from a power law within the mass range of observations, so we wish to determine whether this constitutes evidence that these galaxies have IMFs different from the MW. 

To this end, we draw masses from a Chabrier lognormal IMF and fit them with both a power law and a lognormal IMF, for observations reaching a range of values of $M_{\rm obs}$ and $M_{\rm star}$. Then, for each set of model fits, we use the Bayesian Information Criterion \citep[BIC;][]{schwarz_1978} to evaluate whether the sample of masses is fit significantly better by a lognormal IMF than by a power law. The BIC is a rough diagnostic for comparing how well different models fit a particular dataset. It is designed to balance higher likelihood against increasing model complexity, since more complex models with larger numbers of free parameters can often produce higher likelihood through overfitting.

In general, models with  lower BIC are preferred.  In Figure~\ref{fig:delta_bic}, we plot the difference in BIC between a power law and lognormal IMF as a function of $M_{\rm obs}$ for different values of $M_{\rm star}$. A common rule of thumb is that a $\Delta \rm BIC > 10$ is considered ``decisive'' evidence against the model with higher BIC \citep{Liddle_2007}; thus, IMF realizations which lie above the dashed line in Figure~\ref{fig:delta_bic} strongly favor a lognormal IMF over a power law. Points below the line indicate that there is not a strong preference for one model or the other, or (for $\rm \Delta BIC < -10$), that a power law model is strongly preferred over a lognormal.

Figure~\ref{fig:delta_bic} shows that for $M_{\rm star} = 10^4 M_{\odot}$, the lognormal fit is not decisively preferred until $M_{\rm obs}\sim 0.2$, comparable to $m_c$. This can be understood intuitively: the lognormal IMF has a nearly constant slope at $m \gg m_c$, and it is thus not possible to distinguish between a simple power law and and an IMF which turns over at lower masses with insufficiently deep data. However, statistical power to constrain the IMF increases with increasing $M_{\rm star}$, so that at fixed $M_{\rm obs}$, deviations from a power law IMF can be detected more robustly for larger $M_{\rm star}$. 

We note that none of the published IMF constraints shown in Figure~\ref{fig:alpha_vs_mobs} reach sufficiently low masses to robustly rule out a turnover at the Chabrier characteristic mass. The constraints from \citet{Kalirai_2013} come closest, but we find that for their sample size ($\sim 5000$ stars with masses below $0.93M_{\odot}$), observations would have to reach down to $\sim 0.25 M_{\odot}$ before the BIC would decisively favor a lognormal to a power law IMF.

In conclusion, we have shown that published star count data in the Local Group do not reach sufficiently low stellar masses to robustly measure the low-mass turnover in the IMF: existing studies have only measured the slope of the IMF over a narrow mass range ($0.4M_{\odot} \lesssim m \lesssim 0.8 M_{\odot}$) in which a lognormal IMF is effectively indistinguishable from a power law. If the IMF is lognormal, observations must reach approximately down to $m_c$, which for a Chabrier IMF is $m\approx 0.22$ ($M_{V}\approx 10.2$ for an old, metal-poor stellar population). 

In addition, we find that the shallower IMF slopes reported in Local Group dwarf galaxies thus far between masses of $\sim 0.4 M_{\odot}$ and $\sim 0.8 M_{\odot}$ are approximately consistent with a Chabrier IMF, despite apparent disagreement with the Kroupa slope in this mass regime. We emphasize that deeper observations are required for robust detection of IMF variations. 

\subsection{Fixing model parameters during fitting}
\label{sec:fixing_params}

\begin{figure}
\includegraphics[width=\columnwidth]{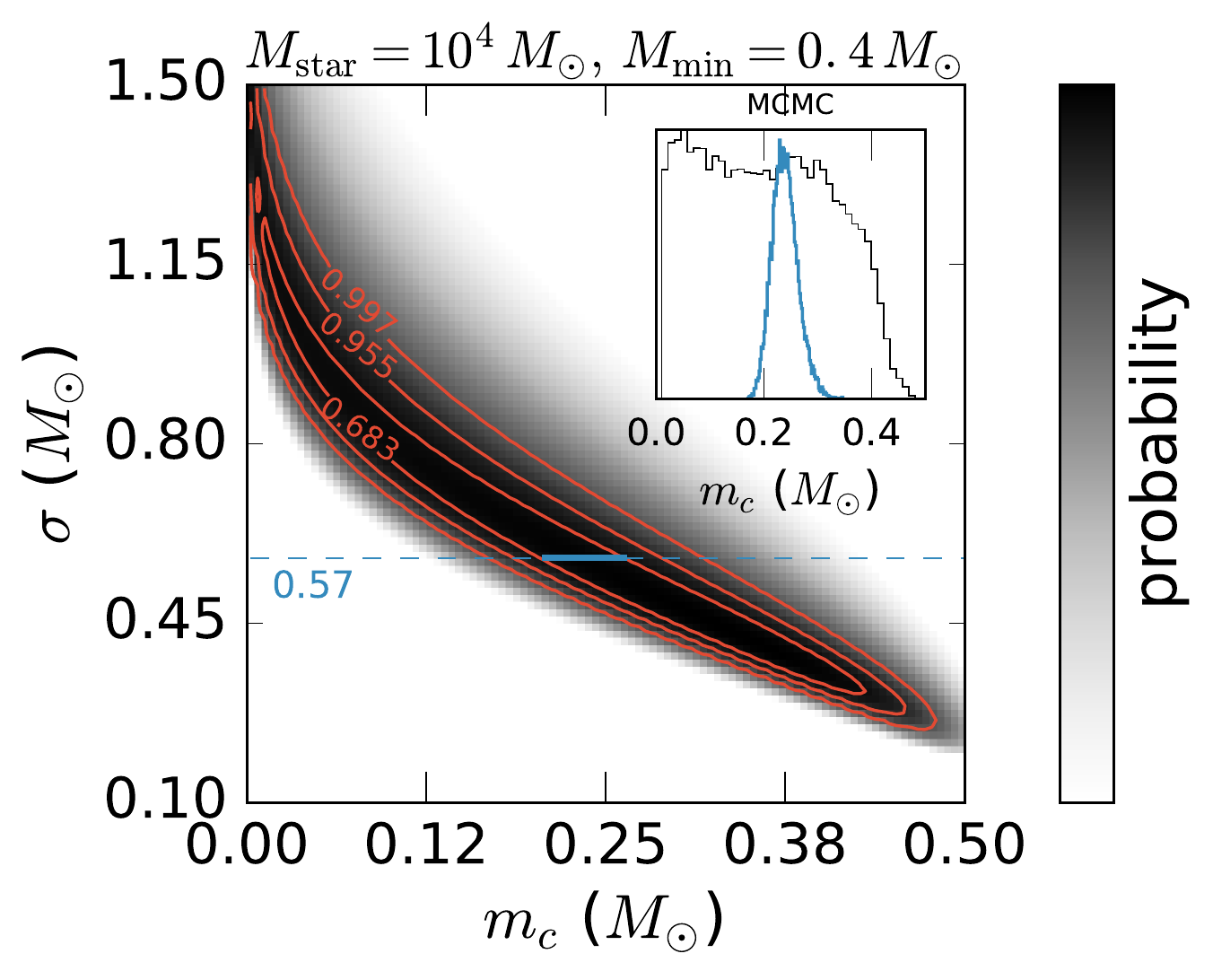}
\caption{Likelihood for fitting a lognormal IMF. Orange contours enclose regions corresponding to 68.3\%, 95.4\%, and 99.7\% probability. Inset shows samples from the posterior distribution for $m_c$ obtained from MCMC fitting. Blue histogram corresponds to holding $\sigma$ fixed;  black histogram shows results of allowing $\sigma$ to vary and marginalizing. Because $m_c$ and $\sigma$ are highly degenerate, holding $\sigma$ fixed during fitting significantly underestimates the uncertainty in $m_c$.}
\label{fig:perils_of_fixing_params}
\end{figure}

Instead of fitting a power law IMF, another possible approach for constraining the IMF when data are only available over a narrow range of masses is to hold one of the IMF parameters fixed during fitting. If either $m_c$ or $\sigma$ is held fixed (and $m_c \ll M_{\rm obs}$, so the IMF resembles a power law), then varying the parameter which is not held fixed effectively varies the power law slope at $m \gg m_c$.  

Figure~\ref{fig:perils_of_fixing_params} shows isoprobability contours for fitting a lognormal IMF to a sample of masses and compares the constraints on $m_c$ obtained when $\sigma$ is held fixed at $\sigma = 0.57 M_{\odot}$ (the Chabrier value) to those obtained when both $\sigma$ and $m_c$ are allowed to vary. In the inset, we show draws from the marginalized posterior for $m_c$ when both parameters are allowed to vary (black) and when $\sigma$ is held fixed at $0.57 M_{\odot}$ (blue). 

When both parameters are left free, $m_c$ and $\sigma$ are highly covariant. Although essentially any $m_c \lesssim 0.4 M_{\odot}$ provides a good fit when paired with an appropriate choice of $\sigma$, all acceptable fits reside in a nearly one-dimensional region of parameter space. Thus, when $\sigma$ is held fixed, fitting a lognormal yields much stronger constraints on $m_c$. 

It is important to note, however, that holding $\sigma$ fixed essentially amounts to putting a delta function prior on $\sigma$, which is generally not justified. Because of the strong degeneracy between $m_c$ and $\sigma$, holding $\sigma$ fixed will dramatically underestimate the true uncertainty in $m_c$: naively interpreting the blue posterior for $m_c$ in Figure~\ref{fig:perils_of_fixing_params}, one might conclude that the characteristic mass $m_c$ is was well-constrained to $0.2 M_{\odot} < m_c < 0.3 M_{\odot}$ when in fact, a much larger range of values can provide equally good fits. This is not necessarily a problem, \textit{if} the constraints from fitting are compared to other studies in the same mass range, since all acceptable fits have basically the same shape across the mass range where data are available (see Figure~\ref{fig:overview}). 

On the other hand, attempts to extrapolate the fitted IMF to lower masses in order to constrain the turnover mass will yield erroneously strong constraints: clearly, if observations only sample a small range of masses at $m\gg m_c$, where the IMF has not yet begun to turn over, they cannot reliably constrain $m_c$ or the turnover mass. We therefore caution against fixing model parameters during fitting. If observations do not reach sufficiently low-mass stars to measure the turnover in the IMF, it is better to fit a simple power law than to fit a more complicated IMF while fixing one or more of the parameters.   

\section{What is needed to detect variation in the IMF?}
\label{sec:true_variation}

\begin{figure*}
\includegraphics[width=\textwidth]{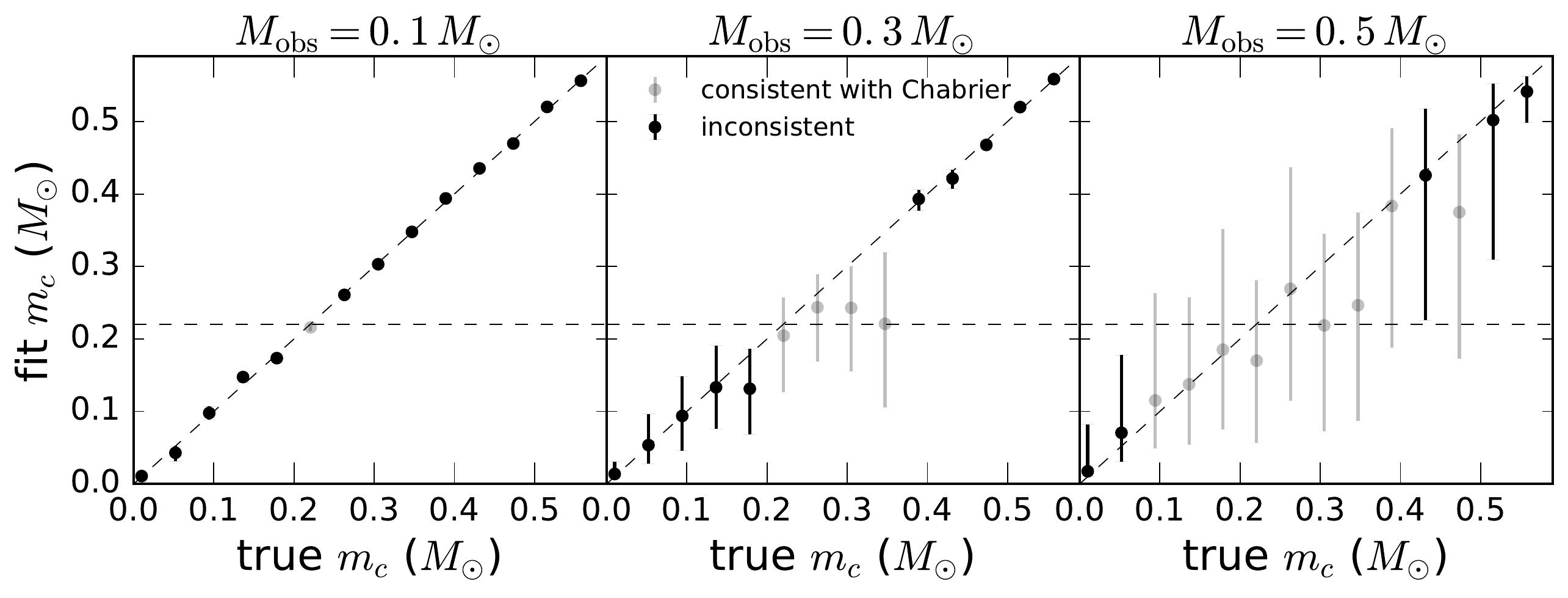}
\caption{Accuracy to which the characteristic mass $m_c$ of a lognormal IMF can be recovered for three different observational stellar mass lower limits, $M_{\rm obs}$. We draw masses from 14 different lognormal IMFs with characteristic masses $m_c$ in the range $0.01 M_{\odot} - 0.6M_{\odot}$, always for a total $M_{\rm star} = 10^4 M_{\odot}$. For each $m_c$, we choose $\sigma$ such that the IMF at masses above $0.5\,M_{\odot}$ resembles a Chabrier IMF (see Figure~\ref{fig:overview}). We then fit a lognormal IMF to the drawn masses. Points and error bars show the median and middle 68\% of the marginalized posterior draws for $m_c$. Gray points indicate that the data cannot rule out a Chabrier IMF with $m_c = 0.22M_{\odot}$; black points indicate that $m_c = 0.22 M_{\odot}$ falls outside the 68\% constraints on $m_c$, and thus, that the data appear to be inconsistent with being drawn from a Chabrier IMF.}
\label{fig:variable_imf}
\end{figure*}

One of the goals of studying the IMFs of dwarf spheroidal and ultra-faint galaxies is to detect or rule out systematic IMF variations with environment. In this section, we investigate how deep observations must reach in order to robustly detect variation in the IMF (if it exists). We assume that the IMF is lognormal, but that the characteristic mass $m_c$ varies across galaxies. Since observations of nearby dwarfs have found only marginal evidence of IMF variations at $m \gtrsim 0.5 M_{\odot}$, we intentionally choose $m_c$ and $\sigma$ so that all IMFs have approximately the same slope at $m\sim 0.5 M_{\odot}$. 

Figure ~\ref{fig:variable_imf} shows how accurately the form of the IMF can be recovered for different $M_{\rm obs}$, the minimum mass down to which observations are complete. For each IMF sample realization, we plot the median and middle 68\% percent range of the marginalized posterior for $m_c$ against the true $m_c$ of the IMF from which the sample was drawn. If the middle 68\% region contains $m_c = 0.22 M_{\odot}$, we label the IMF realization ``consistent with Chabrier''; i.e., observing such a sample of masses would not raise any suspicion of variation in the IMF. If the Chabrier value of $m_c = 0.22M_{\odot}$ falls outside the middle 68\% credibility region, we label the realization inconsistent with the Chabrier value. Of course, if the middle 68\% credibility region extends almost to $m_c = 0.22M_{\odot}$, this would not represent strong evidence of variation in the IMF.\footnote{We also experimented with using a KS test as a non-parametric measure of whether a given sample of masses was consistent with being drawn from a Chabrier IMF, finding results that were similar with the metric described above. We opted to use the parametric version for our primary analysis because we found significant variations in the KS-test $p$ values across different realizations of draws from identical IMFs. This is a generic shortcoming of the KS test; see \citet{Ivezic_2014}.}

Figure~\ref{fig:variable_imf} shows that variation in the IMF can only be robustly detected if $M_{\rm obs}\sim m_c$; that is, if observations are deep enough to detect the beginning of the turnover. In the right panel, where $M_{\rm obs} = 0.5 M_{\odot}$, a wide range of IMFs with $0.1 M_{\odot} \lesssim m_c \lesssim 0.5 M_{\odot}$ are all consistent with the Chabrier value. In other words, if observations only reach down to $0.5 M_{\odot}$, one could observe galaxies with lognormal IMFs and characteristic masses anywhere in the range $0.1 M_{\odot} \lesssim m_c \lesssim 0.5 M_{\odot}$, and the data would not provide sufficiently strong constraints to rule out a Chabrier IMF with $m_c = 0.22 M_{\odot}$. 

However, it becomes much easier to accurately constrain IMF parameters, and thus, to detect deviations from a Chabrier IMF, for lower $M_{\rm obs}$. For $M_{\rm obs} = 0.3 M_{\odot}$ (middle), only the IMFs with $0.2 M_{\odot} \lesssim m_c \lesssim 0.4 M_{\odot}$ are approximately consistent with the Chabrier value, and typical 68\% error bars are smaller than for $M_{\rm obs}=0.5 M_{\odot}$. Finally, for $M_{\rm obs} = 0.1 M_{\odot}$, $m_c$ can be constrained with percent-level accuracy in all cases, and all the IMFs with $m_c$ different from the Chabrier value can be ruled out. The dramatically improved constraints on $m_c$ for lower $M_{\rm obs}$ in this (admittedly idealized) experiment are the result of two factors: first, observations reach low enough that the turnover can be detected directly for lower $M_{\rm obs}$, and second, the total number of stars contributing to the fit increases for deeper observations. 

We note that, even for the cases with higher $M_{\rm obs}$, it is always possible to obtain reasonably tight constraints on $m_c$ (and to detect deviations compared to the Chabrier value) when $M_{\rm obs} \lesssim m_c$. This is because when $M_{\rm obs} \lesssim m_c$, the low-mass turnover is directly probed by observations, and the IMF is no longer well-fit by a power law over the mass range where data are available. Thus, there is no longer a significant degeneracy between $m_c$ and $\sigma$, and both parameters can be constrained strongly. Existing observations can thus already rule out lognormal IMFs with $m_c \gtrsim 0.4 M_{\odot}$.

\section{Looking forward to JWST}
\label{sec:looking_forward}
\begin{figure}
\includegraphics[width = \columnwidth]{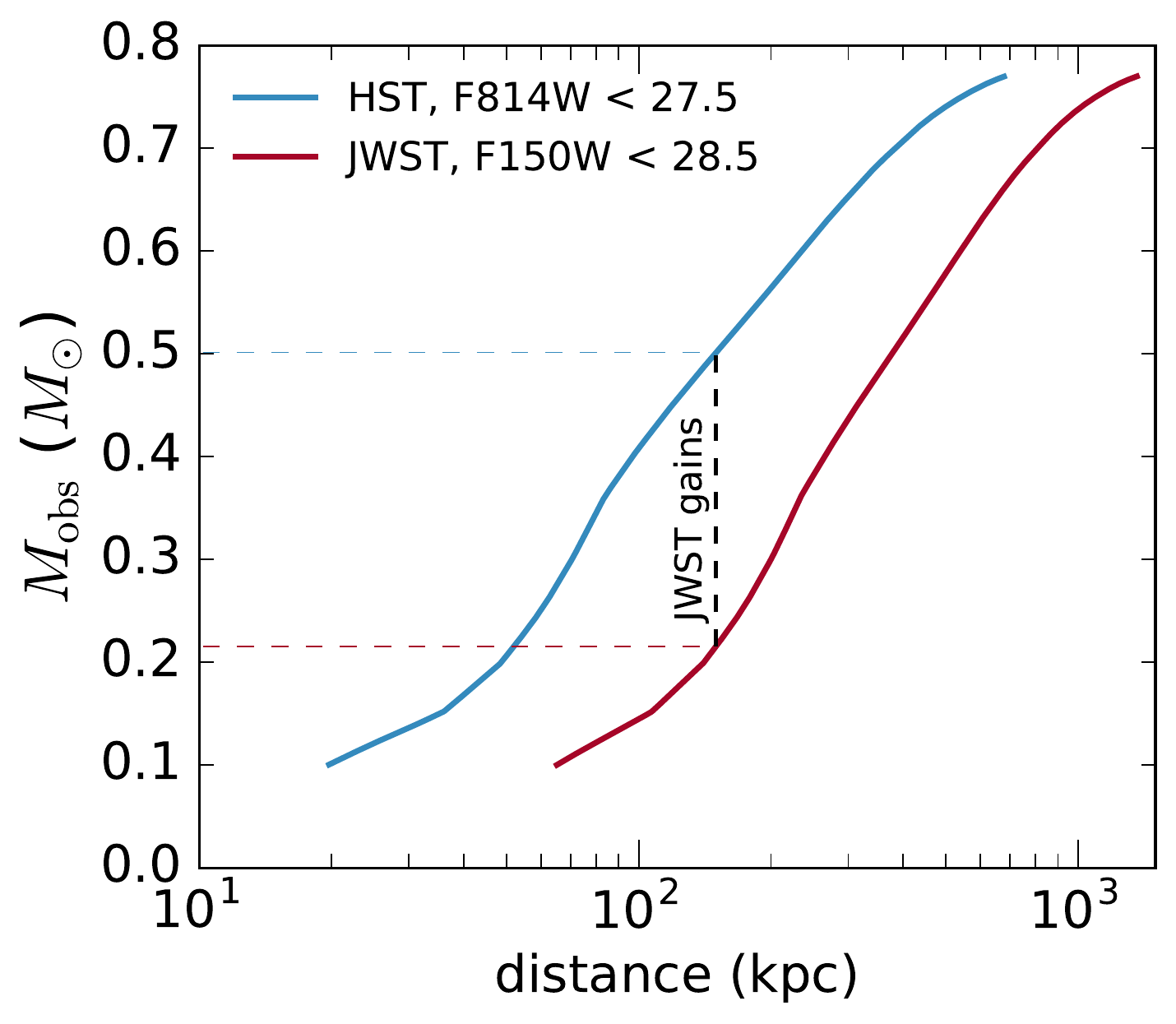}
\caption{Lowest observable stellar mass as a function of distance for HST ACS and JWST observations, assuming 20,000 seconds of integration time. At a fixed distance, JWST can reach stars nearly $0.3 M_{\odot}$ less massive than HST. JWST observations will reach the hydrogen burning limit for distances $\lesssim 60\,\rm kpc$.}
\label{fig:jwst_vs_hst}
\end{figure}

Having quantified the sample size and limiting magnitude required to constrain generic lognormal IMFs using star counts, we now turn to the prospect of measuring the IMFs of Local Group dwarf galaxies with JWST. How accurately the IMF of an individual galaxy can be constrained depends on $M_{\rm obs}$ (which is a function of the magnitude down to which observations are complete) and on the total number of stars sampled (which is a function of the angular size of the galaxy and the number of pointings). 

We begin by calculating a realistic limiting magnitude and corresponding $M_{\rm obs}$ for observations with JWST. Because cool, low-mass stars are most luminous in the NIR (for a typical low-metallicity M-dwarf, $f_{\lambda}$ peaks near $\rm 0.9\mu m$), it is best to use filters near this wavelength. We use the STScI JWST exposure time calculator\footnote{https://jwst.etc.stsci.edu/} to determine the optimal filters for observing low-mass stars and the projected limiting magnitude in these filters. We find that the $F150W$ and $F200W$ filters are optimal for observations of low-mass stars, because although the peak in $f_{\lambda}$ is somewhat bluer, these filters are significantly wider than the $F090W$ and $F115W$ filters and thus admit more total photons.  We find that for an exposure time of approximately 20,000 seconds in the $F150W$ filter, JWST can obtain a signal-to-noise ratio $\rm SNR = 5-10$ or better for stars brighter than 28.5 Vega mag. 

Figure~\ref{fig:jwst_vs_hst} shows the expected lowest mass $M_{\rm obs}$ which JWST will reach as a function of distance. We compute $M_{\rm obs}$ from apparent magnitude at a given distance by interpolating on a grid of isochrones using the \textsc{Mist} code. For comparison, we show the approximate $M_{\rm obs}$ limit of current optical observations with HST.\footnote{Previous studies attempting to constrain the IMFs of nearby dwarf galaxies have used the HST $F606W$ and $F814W$ filters rather than NIR filters, both because HST is significantly less sensitive (and has poorer angular resolution) in the NIR than at optical wavelengths, and because the low-mass stars reached by previous studies ($m\sim 0.5 M_{\odot}$) are somewhat less red than the lowest-mass stars that can be reached with JWST.} This HST curve is computed assuming a magnitude limit of 27.5 Vega mag in the $F814W$ filter, which can be obtained for an exposure time and desired $S/N$ ratio similar to our assumptions for JWST \citep{Geha_2013,Kalirai_2013}. 

Promisingly, JWST is forecasted to reach significantly lower mass stars at fixed distance. For example, at $d = 150$ kpc, the approximate distance of the most-distant MW satellites for which previous studies have attempted to measure the IMF \citep{Geha_2013}, HST observations can only reach down to $M_{\rm obs} \approx 0.5M_{\odot}$ with typical integration times, which, as we showed in Section~\ref{sec:recovering_c03}, is not deep enough  to jointly constrain $\sigma$ and $m_c$ for a Chabrier IMF. At the same distance, JWST is projected to reach down to $M_{\rm obs}\sim 0.22 M_{\odot}$, which is sufficiently deep to detect the low-mass turnover and constrain $m_c$ and $\sigma$ within $\sim 20\%$, if the IMF is similar to a Chabrier IMF. In addition, for the nearest MW satellites at $d \lesssim 60$ kpc, JWST observations will reach stars down to the hydrogen burning limit. These significant improvements over HST stem both from JWST's larger collecting area and from its sensitivity to longer wavelengths, where low-mass stars are more luminous. 

\subsection{Optimal targets in the Local Group}
\label{sec:planning_ahead}

\begin{figure}
\includegraphics[width=\columnwidth]{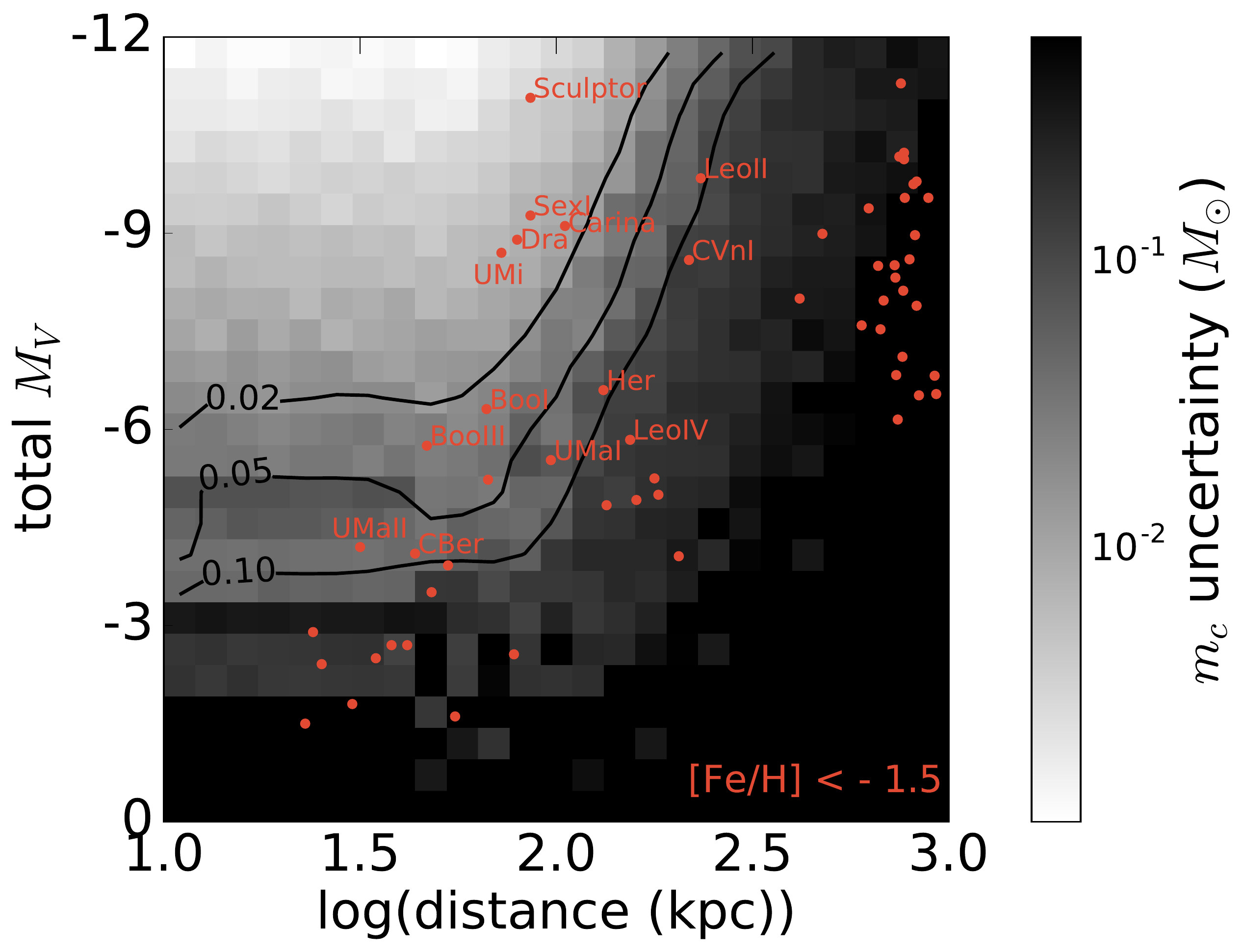}
\caption{Projected accuracy to which the IMF can be recovered from JWST observations of Local Group dwarf galaxies. Gray pixels show the strength of constraints on $m_c$ obtained by fitting a lognormal IMF to samples of stars drawn for populations with different $M_{\rm obs}$ and total $M_{V}$; lighter pixel colors indicate stronger constraints. Local Group dwarf galaxies with $\rm [Fe/H]<-1.5$ are overplotted, with the most promising targets labeled.}
\label{fig:real_galaxies}
\end{figure}

We now combine the projected sensitivity of JWST with the dependence of IMF constraints on $M_{\rm star}$ and $M_{\rm obs}$ derived in the previous sections to devise an observational strategy for constraining the low-mass IMF in the Local Group. Nearer galaxies can be observed down to fainter $M_{\rm obs}$, but there are more massive galaxies, which offer better statistical power (i.e., larger $M_{\rm star}$), at larger distances. An additional consideration is that more distant objects subtend a smaller angular size on the sky, and thus require fewer pointings, but are also more likely to be affected by crowding. 

We investigate the tradeoff between distance and the number of observable stars in Figure~\ref{fig:real_galaxies}, which shows the constraints on $m_c$ expected for a galaxy at a given distance (which determines $M_{\rm obs}$) with a given absolute magnitude (which determines $M_{\rm star}$). For a given total $M_{V}$, the total stellar mass is 

\begin{equation}
\label{eqn:mstar_mv}
M_{{\rm star}}=f_{\rm obs}\Upsilon_{V}\times10^{-M_{V}/2.5}.
\end{equation}
Here $\Upsilon_{V}$ is the $V-$band stellar mass-to-light ratio:\footnote{Note that unlike in the previous sections, the integral is over all stars, including those that have left the main sequence. The resulting $\Upsilon_{V}$ represents a time-averaged quantity, which can fluctuate significantly due to stochastic sampling in the faintest galaxies, whose light is dominated by a small number of evolved stars \citep{Hernandez_2012}.} 

\begin{equation}
\label{eqn:upsilon}
\Upsilon_{V}=\frac{\int_{M_{{\rm min}}}^{M_{{\rm max}}}\xi\left(m\right)m\,{\rm d}m}{\int_{M_{{\rm min}}}^{M_{{\rm max}}}\xi\left(m\right)L_{V}\left(m\right)\,{\rm d}m},
\end{equation}
where $L_{V}(m)$ is the $V-$band luminosity of a star with initial mass $m$, normalized at the magnitude zeropoint. In practice, we calculate $L_{V}(m)$ for a population of a given age and metallicity by interpolating on a grid of stellar models. We assume that observations will only observe a fraction $f_{\rm obs} = 0.1$ of all stars above $M_{\rm obs}$, due to the limited number of pointings and the difficulty of distinguishing between the satellite and foreground MW stars in the galaxy's outskirts. That is, $f_{\rm obs}$ represents the fraction of a galaxy's angular area that is imaged, and the observable $M_{\rm star} = f_{\rm obs} M_{\rm star,tot}$. 

In each pixel of Figure~\ref{fig:real_galaxies}, we calculate the value of $M_{\rm star}$ corresponding to $M_{V}$ using Equation~\ref{eqn:mstar_mv}, and the value of $M_{\rm obs}$ by interpolating the JWST curve in Figure~\ref{fig:jwst_vs_hst}. We then draw masses between $M_{\rm obs}$ and $M_{\rm max}$ from a Chabrier IMF of total mass $M_{\rm star}$ between $M_{\rm min}$ and $M_{\rm max}$, following the procedure outlined in Section~\ref{sec:recovering_true_imf}. Finally, we fit a lognormal IMF to the drawn masses using our standard MCMC procedure outlined in Section~\ref{sec:fitting}. We measure the 68\% uncertainty in the recovered value of $m_c$ and color each pixel in Figure~\ref{fig:real_galaxies} accordingly. Lightly shaded pixels indicate that $m_c$ could be recovered with little uncertainty, while dark pixels correspond to large uncertainties and poor constraints on $m_c$. 

In Figure~\ref{fig:real_galaxies}, we also overplot Local Group dwarf galaxies that represent possible targets for observational studies to measure the IMF from direct star counts. Data for these galaxies are taken from the Nearby Dwarfs Database presented in \citet{McConnachie_2012}. We only plot low-metallicity objects with measured $\rm [Fe/H] < -1.5$. These represent the most ``extreme'' environments, in which one might most expect systematic IMF variations compared to the MW. In addition, these galaxies typically have primarily old stellar populations with simple SFHs, which makes it easier to determine their IMFs from a CMD. ``Promising'' objects (i.e., those in regions of parameter space where our simulations suggest that the IMF could be constrained with little uncertainty) are labeled. Points in the upper right corner are primarily the satellites of M31, which are too faint and crowded to be promising targets. 

So far, we have only considered the $M_{\rm obs}$ and $M_{\rm star}$ that can be achieved for targets of a given $M_V$ at a given distance, without taking into account possible complications due to crowding or the number of pointings required for a given source. We now consider the effects of these issues, in order to refine the list of promising targets.

\begin{figure}
\includegraphics[width=\columnwidth]{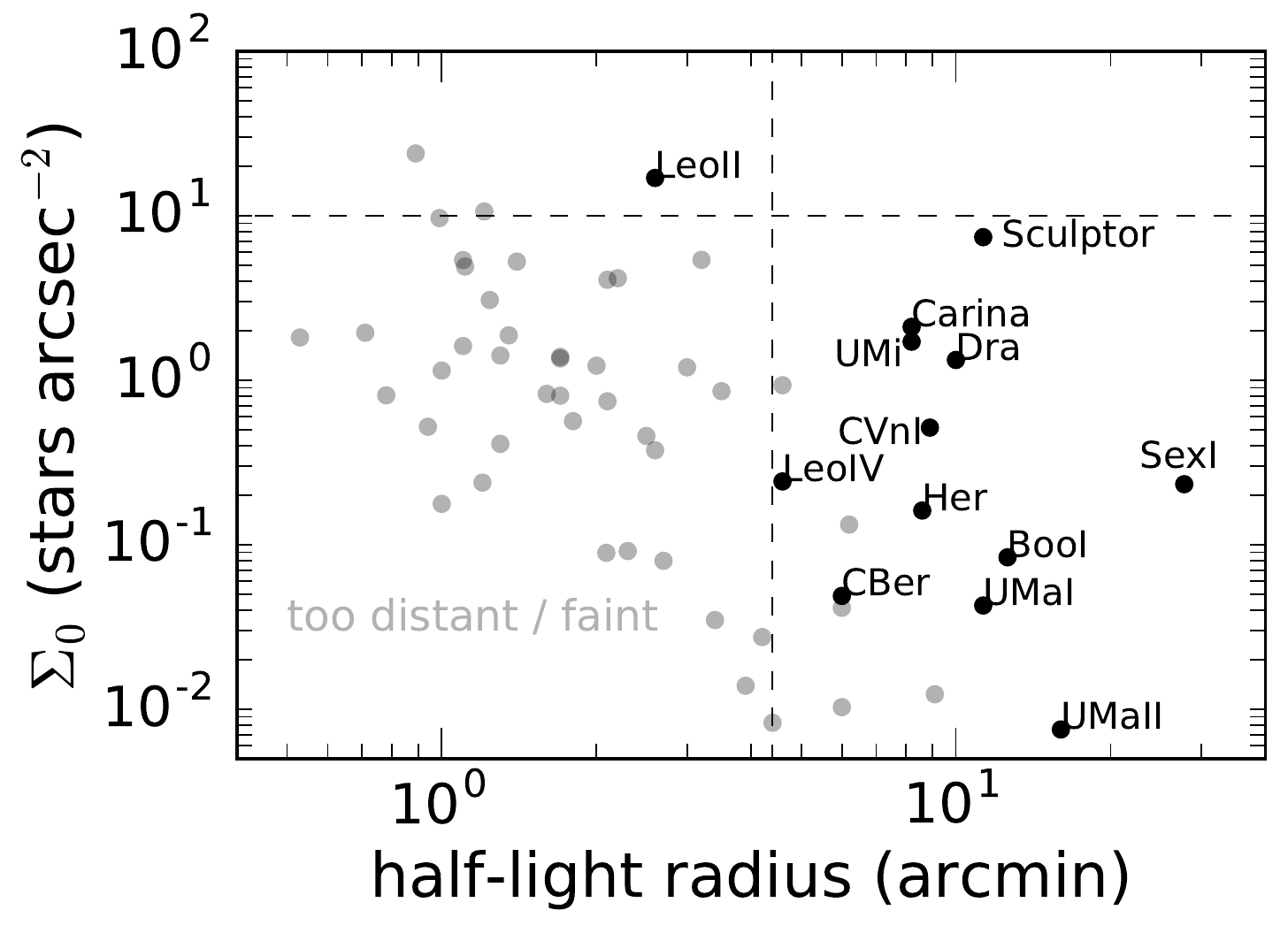}
\caption{Maximum central stellar surface density vs. half-light radius of the same Local Group dwarf galaxies plotted in Figure~\ref{fig:real_galaxies}. Gray dots represent targets found to be unfavorable in Figure~\ref{fig:real_galaxies}, which are also not labeled in Figure~\ref{fig:real_galaxies}. For comparison, we plot as dashed lines the angular size of JWST's NIRCam field of view (4.4 arcmin) and the surface density above which observations will become limited by crowding ($\approx 10 \rm\,stars\,arcsec^{-1}$). }
\label{fig:crowding}
\end{figure}

In Figure~\ref{fig:crowding}, we plot the half-light radius, $r_h$, and maximum central surface number density of stars, $\Sigma_{0}$, of the same galaxies shown in Figure~\ref{fig:real_galaxies}. We roughly estimate $\Sigma_{0}$ from $M_V$, $r_h$, and distance under the assumption that stars follow an exponential surface density profile:
\begin{equation}
\label{eqn:sigma_0}
\Sigma_{0}=\frac{N_{{\rm obs}}}{2\pi a^{2}},
\end{equation}
where $N_{\rm obs}$ is the number of stars above $M_{\rm obs}$ (see Equation ~\ref{eqn:N_obs}), and $a = r_h/1.678$ is the exponential scale length (conversion from \citet{Wolf_2010}). Note that $\Sigma_{0}$ represents the maximum surface density of stars above the $M_{\rm obs}$ reachable with a 20,000 JWST second exposure, not the total surface density of stars down to the hydrogen burning limit. 

We also show the angular size of JWST's NIRCam (4.4 arcmin; 2.2 in the other dimension), and the surface density threshold of 10 stars arcsec$^{-2}$, which is approximately the surface density above which observations become significantly crowding limited. This value is determined under the assumption that JWST's crowding limits will be comparable to those of HST in optical filters, using the crowding data for HST from \citet{Dalcanton_2012}.   

Ideal targets would have both low surface densities (so that crowding is not an issue) and small half-light radii (so that a small number of pointings can image the whole galaxy). Figure~\ref{fig:crowding} shows that crowding should not present a significant problem for most of the promising targets, most of which have maximum stellar surface densities more than a factor of 10 below the assumed crowding threshold. On the other hand, nearly all of the promising targets are significantly larger than the JWST field of view. 

We note that, since we assumed that only a fraction $f_{\rm obs} = 0.1$ of stars would be imaged in creating Figure~\ref{fig:real_galaxies}, it is not necessary to image the entire galaxy to obtain the projected constraints on $m_c$. Nevertheless, the majority of the targets labeled in Figure~\ref{fig:crowding} would require more than one pointing. This is not altogether surprising, since constraining the IMF at low masses requires low $M_{\rm obs}$, which can only be reached for nearby satellites, and nearby satellites subtend the largest angular size at fixed $M_V$. 

One possibility for observations in the post-JWST era is to use WFIRST, which will have a field of view nearly 100 times larger than JWST, and thus could image any of our promising targets in a single pointing \citep{Spergel_2013}. Unfortunately, WFIRST will have significantly poorer angular resolution than JWST, and will thus be more crowding limited. In addition, WFIRST will not reach as low an $M_{\rm obs}$ as JWST at fixed distance and exposure time due to its smaller collecting area. Thus, WFIRST will be ideally suited for measuring the IMFs of the very nearest MW satellites (e.g. Coma Berenices, Ursa Major I, and Bo{\"o}tes I), which are near enough to allow deep observations but subtend a large angular area and would require a large number of pointings with JWST or HST.

\section{Summary}
\label{sec:summary}
Using simple Monte Carlo simulations, we have explored the statistical challenges of constraining the sub-stellar IMF in Local Group dwarf galaxies with direct star counts. We consider both the idealized case in which stellar masses are known exactly and the IMF can be fit directly, and the more realistic case in which the IMF is determined by comparing model CMDs to photometry (see Appendix~\ref{sec:cmd_fitting}). Our main results are as follows. 

\begin{enumerate}[leftmargin=*]
\item \textit{Statistical constraints on the turnover in the IMF}: In order to break the degeneracy between the characteristic mass ($m_c$) and width ($\sigma$) when fitting a lognormal IMF, observations must reach a minimum stellar mass $M_{\rm obs} \sim m_c$; i.e., approximately $M_{\rm obs} = 0.2 M_{\odot}$ if the IMF is similar to the Chabrier IMF measured in the MW (see Figure~\ref{fig:sample_size_and_limiting_mag}). Stronger constraints can be obtained for shallower observations, but this typically requires a prohibitively large sample of stars. Existing HST observations have not yet reached sufficiently low masses to jointly constrain $m_c$ and $\sigma$. 

\item \textit{Fitting a power law to masses drawn from a lognormal IMF yields shallow power law slopes}: The degeneracies of fitting a lognormal IMF can be avoided by fitting a single power law $\xi(m) \propto m^{-\alpha}$. However, if observations can only sample the IMF in a narrow mass range such as $0.5 M_{\odot} \lesssim m \lesssim 0.8 M_{\odot}$, which is typical for attempts to constrain the IMFs of MW satellites with HST, this will yield a power law slope $\alpha \sim 1.55$, even when masses are drawn from a Chabrier IMF (see Figure~\ref{fig:alpha_vs_mobs}). This slope is different from the canonical slope measured in the MW ($\alpha = 2.3$ for a Kroupa IMF, or $\alpha = 2.35$ for a Salpeter IMF), simply because of the mass range probed by observations. Existing reports of systematically shallower IMFs in MW satellite galaxies may thus in fact be consistent with a Chabrier IMF. 

\item \textit{Fixing IMF parameters during fitting can lead to underestimated uncertainty}: A common approximation in fitting a lognormal IMF is to fix the parameter $\sigma$ to e.g. the Chabrier value, in order to avoid the degeneracy with $m_c$. However, this practice should be avoided, as it will dramatically underestimate the uncertainty in $m_c$ (see Figure~\ref{fig:perils_of_fixing_params}). This is not necessarily a problem if the obtained constraints on $m_c$ are only compared to other constraints obtained in the same mass range with the same fixed $\sigma$ value, but it can leading to erroneous indications of IMF variations if values of $m_c$ are compared across different studies. 

\item \textit{Observations required to detect variability in the low-mass IMF}: If the slope of the subsolar IMFs of different galaxies are similar at $m > 0.5 M_{\odot}$ (as appears to be the case), observations must reach down to at least $m_c$ in order to detect variability in the IMF and rule out the possibility that observed data are drawn from a Chabrier IMF (see Figure~\ref{fig:variable_imf}). In addition, failure to detect a turnover in the IMF does not imply deviations from the Chabrier IMF unless observations reach down to $m\sim 0.2 M_{\odot}$ (see Figure~\ref{fig:delta_bic}). 

\item \textit{Prospects for constraining the IMFs of Local Group dwarf galaxies with JWST}: Near-infrared observations with JWST will probe significantly lower stellar masses at fixed distance than previous studies with HST (see Figure~\ref{fig:jwst_vs_hst}), which will allow for unprecedented measurement of the low mass turnover (if it exists) in the IMFs of nearby dwarf galaxies. JWST observations will be complete down to the hydrogen burning limit for galaxies within $\sim 60$ kpc, and down to $\sim 0.22 M_{\odot}$ for galaxies at distances of $\sim 150$ kpc. Promising targets for IMF studies in the Local Group are highlighted in Figure~\ref{fig:real_galaxies}. JWST observations of most nearby dwarf galaxies will not be significantly limited by crowding, but a similar number of pointings will be required per galaxy as with HST observations to build a sufficiently large sample of stars (see Figure~\ref{fig:crowding}).

\end{enumerate}

\section*{Acknowledgements}
We thank the anonymous referee for a constructive report which improved this paper.
We thank Marla Geha and Clement Wagner for comments on the manuscript, and Hans-Walter Rix and Kevin Covey for helpful discussions. 
KE gratefully acknowledges support from a Berkeley graduate fellowship, a Hellman award for graduate study, and an NSF Graduate Research Fellowship.
EQ was supported by NASA ATP grant 12-ATP-120183, a Simons Investigator award from the Simons Foundation, and the David and Lucile Packard Foundation. 
We ran numerical calculations on the Caltech compute cluster ``Zwicky'' (NSF MRI award \#PHY-0960291). This research made use of Astropy, a community-developed core Python package for Astronomy \citep{Astropy_2013}, and Matplotlib \citep{Hunter:2007}. 

\bibliographystyle{mnras}

\appendix
\section{Drawing masses from the IMF}
\label{sec:drawing_masses}
Masses are drawn from the IMF using inverse transform sampling \citep[e.g][]{Olver_2013}. Let $\xi(m)$ be an arbitrary IMF, normalized such that $\int_{M_{{\rm min}}}^{M_{{\rm max}}}\xi\left(m\right)\,{\rm d}m=1$. 
We define the cumulative initial mass function $N\left(m\right)=\int_{M_{{\rm min}}}^{m}\xi\left(m'\right)\,{\rm d}m'$. Then $N(m)$ gives the probability that a random draw from the IMF will be less than $m$. By definition, if  $m_i$ are masses drawn from $\xi(m)$, then $N(m_i)$ will be uniformly distributed; that is, $N(m_i)=\mathcal{U}(0, 1).$

We define the inverse cumulative mass function $N^{-1}(x)$, which takes a number $x\in[0,1]$ and returns the mass $m$ for which the probability of drawing a mass less than $m$ from $\xi(m)$ is $x$. Thus, if $u_i$ are random draws from a uniform distribution $\mathcal{U}(0,1)$, then $N^{-1}(u_i)=N^{-1}(N(m_i))=m_i$  will be random draws from $\xi(m)$. Therefore, to draw masses from $\xi(m)$, one can draw values from $\mathcal{U}(0,1)$ and pass them through $N^{-1}(x)$. 

For a lognormal IMF in the form of Equation~\ref{eqn:chab_imf}, the inverse of the cumulative distribution is given by 

\begin{equation}
\label{eqn:lognorm_inv_cum}
N^{-1}(x)=10^{\left(\sqrt{2}\sigma{\rm erf^{-1}}\left[g(M_{{\rm max}})x+(1-x)g(M_{{\rm min}})\right]+\log m_c \right)},
\end{equation}
where $g(y)={\rm erf}\left[(\log y-\log m_c)/(\sqrt{2}\sigma)\right]$ and ${\rm erf^{-1}}(x)$ is the inverse error function. 

\section{Translating between masses and magnitudes}
\label{sec:stellar_models}

\begin{figure}
\includegraphics[width=\columnwidth]{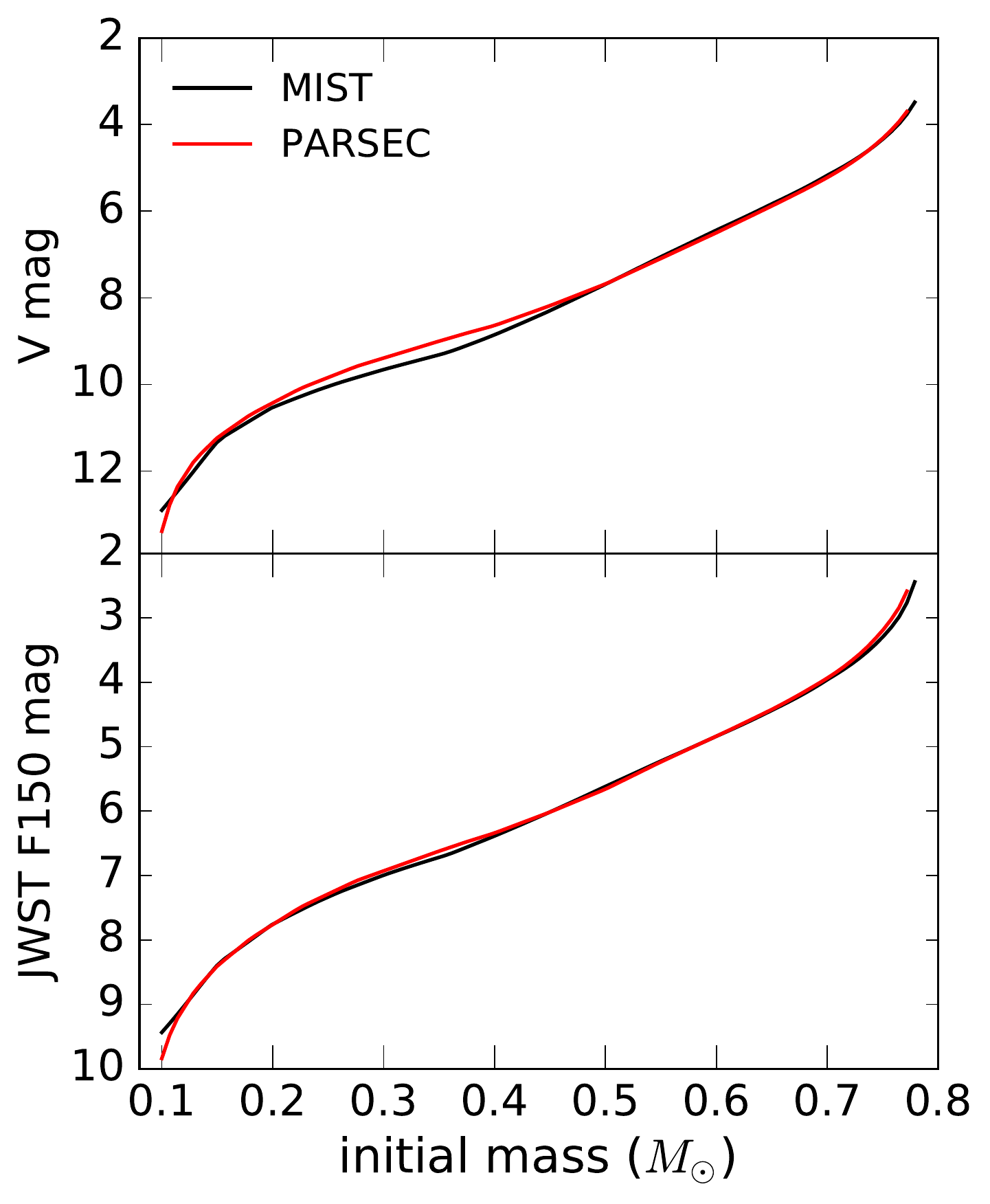}
\caption{Absolute magnitude in the $V$ (top) and JWST $F150$ (bottom) bands, as predicted by two different stellar evolution codes. Both models are  for stars of age 12.5 Gyr and metallicity $\rm [Fe/H]=-2$. We use the MIST models for our primary analysis.}
\label{fig:stellar_models}
\end{figure}

We convert between stellar masses and absolute magnitudes by interpolating on a grid of stellar models. For our primary analysis, we assume a single stellar population with age 12.5 Gyr and metallicity $\rm [Fe/H]=-2$. For a particular stellar evolutionary model, this uniquely specifies the transformation between stellar initial mass and absolute magnitude. We do not attempt to account for complications due to $\alpha-$element enhancement, stellar rotation, or binary evolution, which for our purposes represent small second order corrections.

Figure~\ref{fig:stellar_models} shows the relation between initial mass and absolute magnitude predicted by stellar evolution models. We compare the relations predicted by two different codes: \textsc{Mist} \citep{Choi_2016,Dotter_2016}, which is based on the \textsc{Mesa} code \citep{Paxton_2011,Paxton_2013,Paxton_2015}, and \textsc{Parsec} \citep{Bressan_2012}, which is based on the Padova isochrones \citep{Marigo_2008}. We plot mass-luminosity relations in both the $V$ (top) and the JWST $F150W$ (bottom) bands. 

The two models are, for the most part, in good agreement. The most significant deviation occurs in the $V$ band for stars with initial masses in the range $(0.3-0.4)M_{\odot}$, where the two models disagree by up to $10\%$ on the initial mass corresponding to a given absolute magnitude. This disagreement is likely a result of the sensitivity of the cool atmospheres of low-mass stars to molecular opacity \citep{Allard_1997}. The two models are in better agreement at the same mass in the redder $F150W$ band, which is less strongly affected by molecular absorption in stellar atmospheres.

We emphasize that our main results are all derived by fitting masses directly, without transforming into magnitude space. Uncertainties in stellar models thus do not directly affect our calculations of the IMF constraints that can be obtained for a given $M_{\rm obs}$ and $M_{\rm star}$; they only come into play when these values are translated into magnitudes. On the other hand, if fitting is done directly in magnitude space as in the next section, errors in stellar models could introduce biases in the recovered IMF. 

\section{Extracting the IMF from a CMD}
\label{sec:cmd_fitting}

In this section, we summarize how the IMF can be recovered from a CMD if the  masses of individual stars are not known exactly. We first generate a ``data'' CMD by assigning magnitudes to masses sampled from the IMF and applying a completeness function. We then generate ``model'' CMDs by forward modeling IMFs with different parameters, choosing the best-fit IMF as the one for which the model CMD most closely matches the data CMD.  Our goal is primarily to demonstrate that this approach, which is commonly used in observational studies, does not introduce systematic errors compared to our fiducial method of fitting masses directly, which was described in Section~\ref{sec:fitting}. 

We assume stellar magnitudes have been measured in two filters, which we label $V$ and $I$. Of course, this analysis is in principle applicable to CMDs generated in any two filters. We calculate $V$ and $I$ magnitudes for stars of a given initial mass, metallicity, and age by interpolating on a grid of \textsc{Mist} stellar models. In order to minimize the weight given to stars in poorly understood phases of stellar evolution (which are also not sensitive to the IMF), we include only main-sequence stars in our fit.

\subsection{Stellar metallicity distribution function}
\label{sec:MDF}

\begin{figure}
\includegraphics[width=\columnwidth]{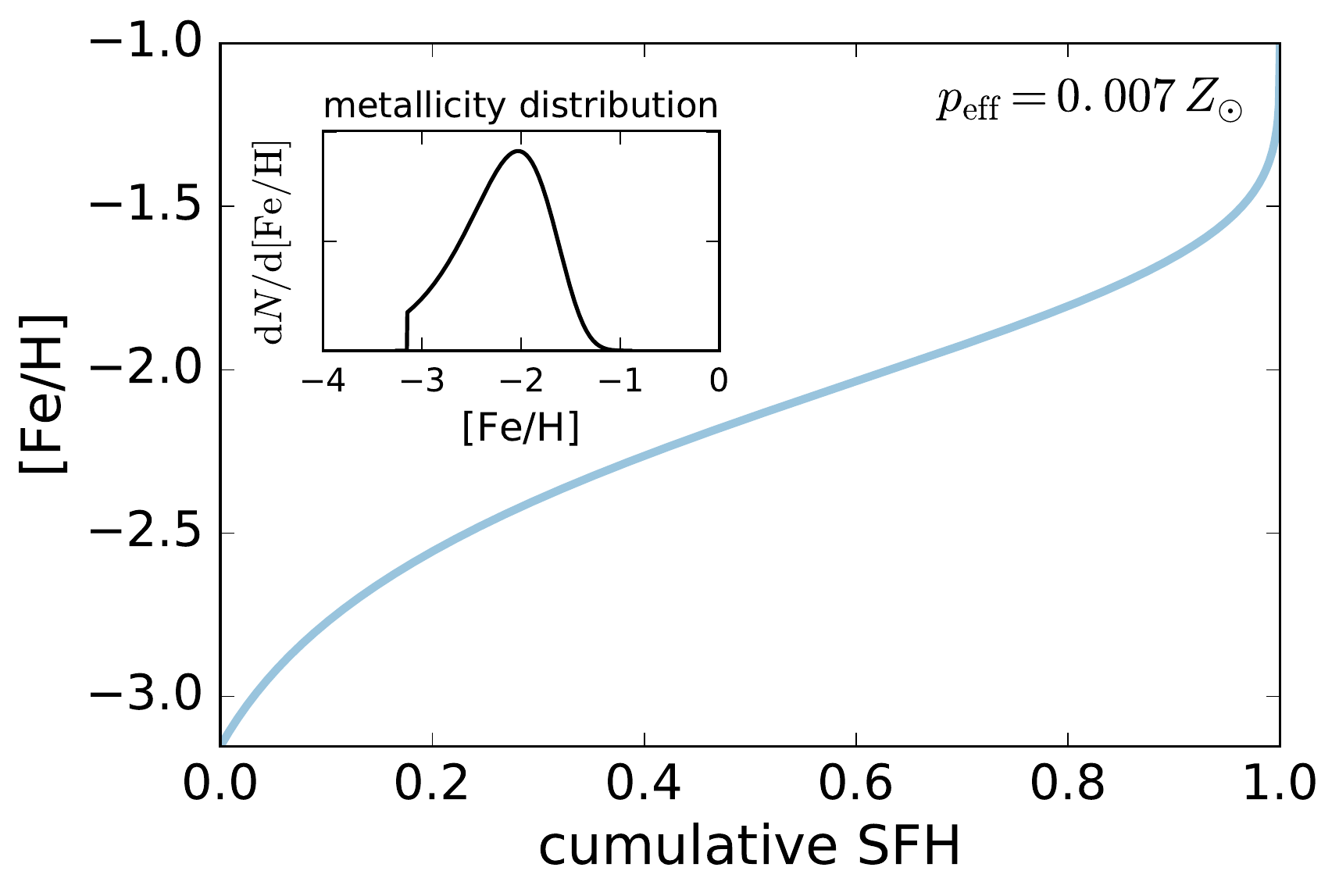}
\caption{Pre-enriched ``leaky box'' chemical evolution model assumed in creating the CMD in Figure~\ref{fig:synthetic_cmd}. This model predicts a metallicity distribution function consistent with observations of many nearby low-mass galaxies.}
\label{fig:leaky_box}
\end{figure}

Since stars of a given initial mass will have different magnitudes and colors at different metallicities, it is important to realistically model the metallicity distribution function (MDF) when generating a CMD. Nearby ultra-faint and dwarf spheroidal galaxies exhibit a significant range of stellar metallicities, even in galaxies with simple star formation histories dominated by old ($>12$ Gyr) stars. Observations of dwarf galaxies in the Local Group \citep[e.g.][]{Simon_2007, Kirby_2008, Kirby_2011, Lai_2011,  Kirby_2013, Ross_2015, Frebel_2016} generally find a mean metallicity of $\rm [Fe/H]\approx -2$, with a range of metallicities as large as $\sim 2\,\rm dex$ (i.e., $\rm -3 \lesssim [Fe/H] \lesssim -1$ for stars in a single galaxy). 

We use the MDF predicted by a ``leaky box'' analytic chemical evolution model \citep{Hartwick_1976, Pagel_1997} to generate our CMDs. We modify the model slightly to allow for pre-enrichment, so that the most metal-poor stars have metallicity $Z_0$ (where we set $Z_0 = 10^{-3}Z_{\odot}$) rather than zero metallicity. The MDF predicted by this model has been shown to be in good agreement with the observed MDFs of many Local Group dwarf spheroidal and ultra-faint galaxies \citep{Kirby_2011, Kirby_2013}. It is parameterized by a mass-loading factor $\eta$, which represents the gas mass lost to outflows per unit mass of stars formed, and the nucleosynthetic yield $p$, which represents the mass of metals returned to the ISM per unit mass of stars formed. $p$ depends only on the IMF and the details of nuclear burning, whereas $\eta$ depends on a galaxy's potential depth and star formation history. 

If the highest metallicity star formed by $z=0$ has metallicity $Z=Z_1$, the fraction of stars with metallicity less than $Z$ is given by 
\begin{equation}
\label{eqn:leaky_box_mass}
\frac{M_{\star}\left(<Z\right)}{M_{\star}\left(<Z_{1}\right)}=\frac{1-\exp\left[\left(Z_{0}-Z\right)/p_{{\rm eff}}\right]}{1-\exp\left[\left(Z_{0}-Z_{1}\right)/p_{{\rm eff}}\right]},
\end{equation}
where $p_{\rm eff} = p/(1+\eta)$ is the effective yield. The corresponding MDF is 
\begin{equation}
\label{eqn:leaky_box_mdf}
\frac{{\rm d}N}{{\rm d\left[Fe/H\right]}}\propto10^{\left[{\rm Fe/H}\right]}\exp\left[Z_{\odot}\frac{10^{\left[{\rm Fe/H}\right]_{0}}-10^{\left[{\rm Fe/H}\right]}}{p_{{\rm eff}}}\right],
\end{equation}
in the region $Z_0 \leq Z \leq Z_1$, and 0 outside this region. Here we have defined $\rm [Fe/H]_0 = \log(Z_0/Z_{\odot}).$

Figure ~\ref{fig:leaky_box} shows the enrichment history and metallicity distribution predicted for $p=0.01$ and $\eta=100$. These values were chosen to produce a mean metallicity $\overline{Z} = 0.01 Z_{\odot}$, comparable to what is found in observed galaxies. We use these values and the MDF shown in Figure~\ref{fig:leaky_box} to generate our CMDs.

\subsection{Completeness}
Thus far, we have assumed that observations can resolve all stars with masses above $M_{\rm obs}$ and none of the stars with masses below $M_{\rm obs}$. In real observations, this limit is manifest as a gradual drop-off in completeness with increasing magnitude, not as a sharp cutoff at a particular stellar mass.

We model this drop-off with a completeness function $c(M_{V})$, which represents the fraction of stars with magnitude $M_V$ (which can, in principle, be measured in any filter) that can be observed. We use the completeness function 
\begin{equation}
\label{eqn:completeness}
c\left(M_{V}\right)=\frac{1}{2}\left[1-\tanh\left(M_{V}-M_{V,0.5}\right)\right].
\end{equation}
Here, $M_{V,0.5}$ is the magnitude at which completeness falls to $50\%$, which is chosen for a given $M_{\rm obs}$ by interpolating on a grid of stellar models.  We use this function because it provides a reasonably good approximation for the empirically-estimated completeness functions found in observational IMF studies of low-mass galaxies \citep{Wyse_2002,Kalirai_2013}. The function for $M_{\rm obs} = 0.2M_{\odot}$ is shown in the inset of Figure~\ref{fig:synthetic_cmd}.  

\subsection{Fitting the CMD}

\begin{figure}
\includegraphics[width=\columnwidth]{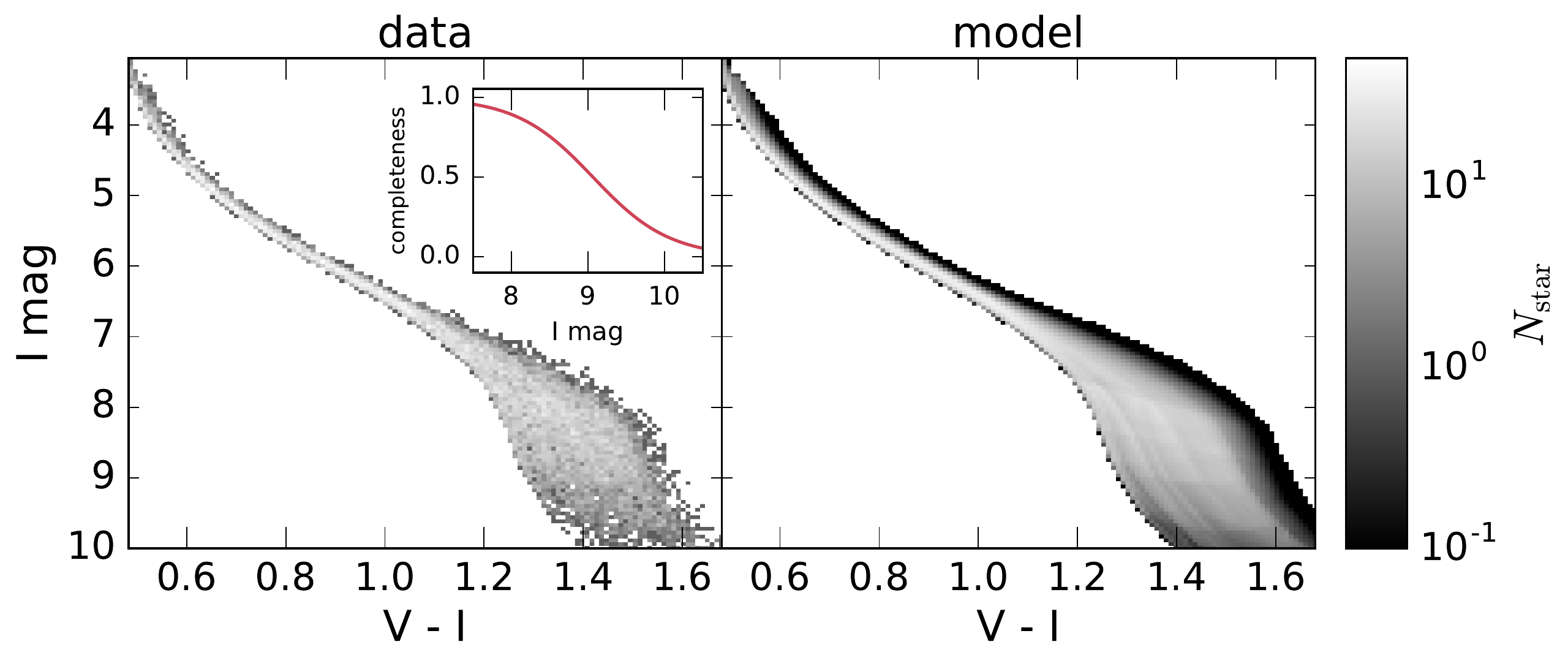}
\caption{\textbf{Left}: synthetic CMD generated for a stellar population with $M_{\rm obs} = 0.2M_{\odot}$ and $M_{\rm star} = 10^4 M_{\odot}$. Only main sequence stars are shown. \textbf{Right}: best-fit model CMD obtained by maximizing the likelihood in Equation~\ref{eqn:poisson_likelihood}. Both the model and the synthetic CMD assume a stellar population of age 12.5 Gyr with a realistic pre-enriched leaky box metallicity distribution function. Inset shows the adopted $I$-band completeness function.}
\label{fig:synthetic_cmd}
\end{figure}

\begin{figure}
\includegraphics[width=\columnwidth]{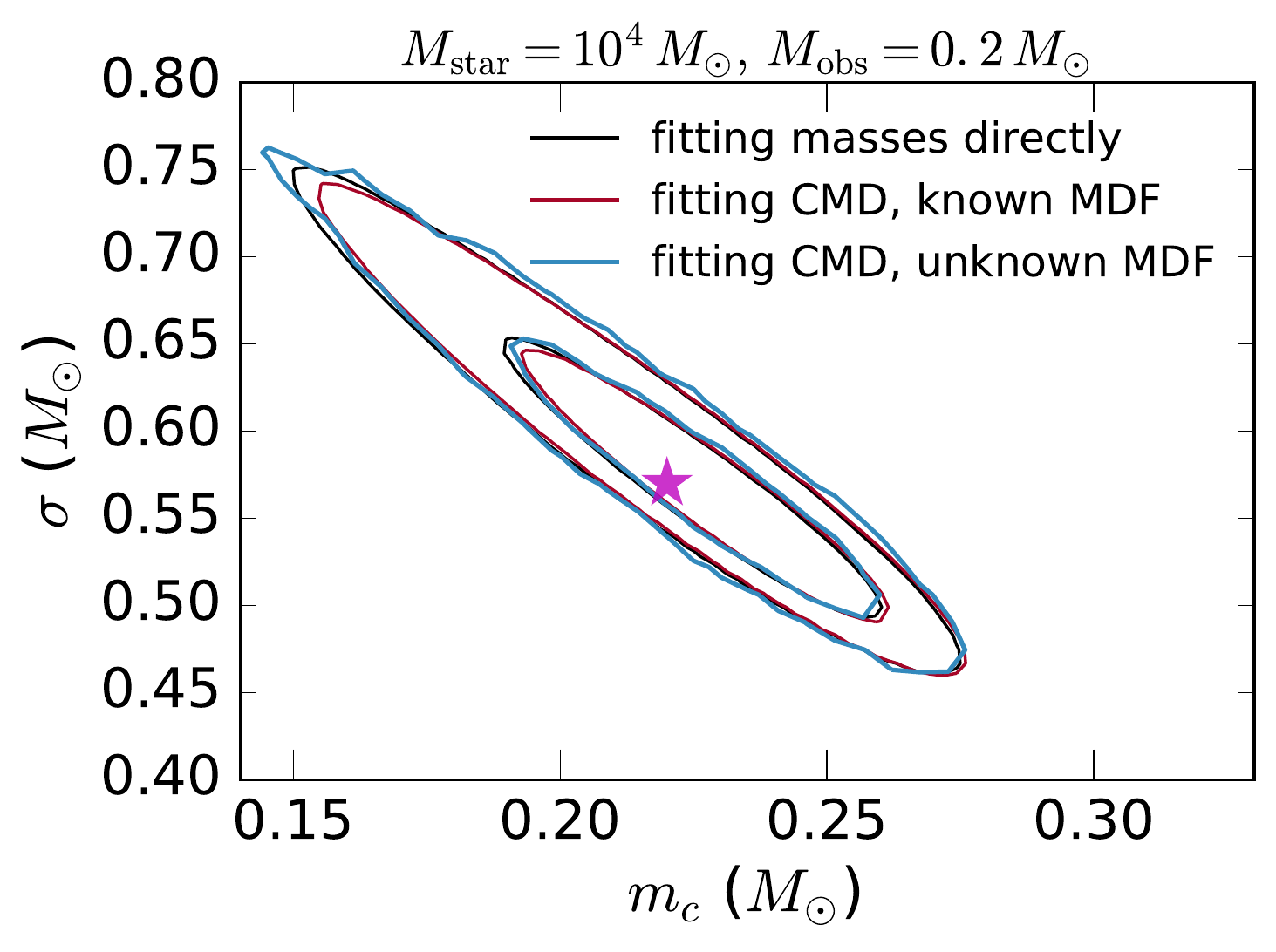}
\caption{Comparison of IMF fits obtained for an example stellar population. Contours show 1 and 2$\sigma$ constraints on the parameters of the IMF obtained with different fitting methods. \textbf{Black}: we fit masses drawn from the IMF directly, as in Equation~\ref{eqn:likelihood}. \textbf{Red}: We generate a synthetic CMD with a known leaky-box metallicity distribution function (MDF) and fit using Equation~\ref{eqn:poisson_likelihood}. \textbf{Blue}: We again fit the synthetic CMD, but now the parameters of the MDF are unknown; contours are generated by marginalizing over the MDF parameters. Star shows the true parameters of the IMF from which masses are drawn.}
\label{fig:cmd_vs_masses}
\end{figure}

We generate a ``data'' CMD by sampling masses from the IMF and metallicities from the MDF, transforming into magnitude space using the \textsc{Mist} stellar models, and discarding stars of magnitude $M_V$ with probability $1-c(M_V)$.

Following the method outlined in \citet{Dolphin_2002}, we then divide the CMD into $N$ bins in both $I$ magnitude and $V-I$ color, so that the full pixelated CMD (or ``Hess diagram'') contains $N^2$ total pixels. We choose $N = \sqrt{N_{\rm star}}$, so that the number of pixels is equal to the number of stars. We have verified that our fit is not sensitive to the choice of bins, as long as bins are small enough to capture the smallest CMD features. 

We generate model CMDs for a population with a given IMF by linearly combing isochrones of different metallicities, weighting points along each isochrone by the IMF and weighting different isochrones by the MDF. We calculate the likelihood for a given model by summing over all pixels, where the likelihood in a given pixel is set by Poisson statistics. Explicitly, if $m_i$ represents the $i-$th pixel of a given model CMD and $d_i$ represents the same pixel in the data CMD, the log-likelihood can be written as 
\begin{equation}
\label{eqn:poisson_likelihood}
\ln\mathcal{L}=\sum_{m_{i}\neq0}\left[d_{i}\ln m_{i}-m_{i}-\ln\left(d_{i}!\right)\right].
\end{equation}
The sum is over all pixels in which the model probability is non-zero. For pixels with $d_i>100$, where it is computationally impractical to evaluate the factorial exactly, we use Stirling's approximation to evaluate $\ln(d_i !)$. For $d_i = 100$, this produces a negligible error of a few $\times 10^{-4}$ percent; the fractional error declines further with increasing $d_i$. We note that while $d_i$ must be a non-negative integer, $m_i$ need not be an integer. 

In Figure~\ref{fig:synthetic_cmd}, we compare a synthetic ``data'' CMD generated for masses drawn from a Chabrier IMF in the mass range $m = (0.2 - 0.77) M_{\odot}$ to the best-fit model CMD generated by maximizing the likelihood in Equation~\ref{eqn:poisson_likelihood}. The model essentially recovers the ``data'' CMD exactly: the apparent differences in the regions where pixels in the model are dark are due to the fact that the model predicts significantly less than 1 star per bin, while the same pixels in the data CMD must have an integer number of stars (typically 0 or 1). 

Figure ~\ref{fig:cmd_vs_masses} shows the probability contours for $m_c$ and $\sigma$ generated by this fitting procedure for a sample of masses drawn from a Chabrier IMF in the mass range $m = (0.2 - 0.77) M_{\odot}$. $V$ and $I-$ band magnitudes were generated assuming a uniform age of 12.5 Gyr and a pre-enriched leaky box MDF. We considered two cases: one in which the parameters of the MDF are known a priori, and one in which we simultaneously fit for the parameters of the MDF and the IMF. In the latter case, the plotted IMF probability contours represent the results of marginalizing over the MDF parameters.  

The constraints on $m_c$ and $\sigma$ obtained by fitting the synthetic CMD are in excellent agreement with those obtained by fitting the IMF directly to the masses as described in Section ~\ref{sec:fitting}. This remains true even when the parameters of the MDF are not known a priori and the MDF is fit simultaneously with the IMF. In other words, there is no loss of information when fitting in color-magnitude space compared to fitting masses directly, even though initial mass does not uniquely determine the magnitude of an individual star for a population with a spread of metallicities. This is because the CMD contains information about the distribution of stellar metallicities: the color distribution of the main sequence is a direct transformation of the MDF. 

This suggests that fitting a CMD provides equally strong constraints on the IMF as fitting masses directly --- in the limit of perfect observations. It is important to remember, however, that this exercise is still highly idealized. In practice, the precision of IMF constraints from CMD fitting could also be limited by photometric uncertainties \citep[see][]{Weisz_2013}, by uncertainties in the stellar models used to translate between mass and magnitude, and by degeneracies between the binary fraction and the MDF. As we have not attempted to model these uncertainties, our IMF constraints represent best-case upper limits.

\end{document}